\documentclass[12pt]{article}

\usepackage{amsmath}
\usepackage{amssymb}
\usepackage{physics}
\usepackage{bm}
\usepackage{comment}

\usepackage{tikz,tikz-cd}
\usepackage{here}

\newcommand{\rev}[1]{\mathrm{rev}(#1)}

\makeatletter
	
	\@addtoreset{equation}{section}

	\@addtoreset{figure}{section}

	\@addtoreset{table}{section}
\makeatother

\begin{document}

\baselineskip=17pt

\begin{titlepage}
\rightline{\tt KUNS-3045}

\begin{center}
\vskip 2.5cm
{\Large \bf {BSFT-like action
from cohomomorphism
}}
\vskip 1.0cm
{\large Jojiro Totsuka-Yoshinaka}
\vskip 1.0cm

{\it {Department of Physics, Kyoto University,}}\\
{\it {Kitashirakawa, Kyoto 606-8502, Japan}}\\
george.yoshinaka@gauge.scphys.kyoto-u.ac.jp,\\
\vspace{2mm}

\vskip 1.0cm

{\bf Abstract}
\end{center}
\noindent
By performing a field redefinition via a cohomomorphism on the Maurer–Cartan action of the $A_\infty$ algebra, we construct an action that reproduces the characteristic structure of boundary string field theory (BSFT), which we refer to as the BSFT-like action. 
We show that the form and several properties of this BSFT-like action closely resemble those of BSFT. 
Our results suggest that this approach provides a systematic framework to derive the BSFT action from the OSFT based on $A_\infty$ algebras.

\end{titlepage}

\tableofcontents

\newpage
\section{Introduction}

Homotopy algebras, such as $A_\infty$ algebras \cite{Stasheff:I,Stasheff:II,Getzler-Jones,Markl,Penkava:1994mu,Gaberdiel:1997ia} and $L_\infty$ algebras \cite{Zwiebach:1992ie,Markl:1997bj}, play a fundamental role in string field theory and quantum field theory. 
Constructing actions based on homotopy algebras not only ensures manifest gauge invariance but also serves as a guiding principle in formulating unknown actions, particularly in string field theory \cite{Maccaferri:2022yzy,Maccaferri:2023gcg,Kajiura:2004xu,Kajiura:2006mt,Erler:2013xta,Erler:2016ybs,Kunitomo:2019glq,Kunitomo:2022qqp}.
Using these algebraic structures enables the derivation of effective actions \cite{Erbin:2020eyc,Koyama:2020qfb,Arvanitakis:2020rrk,Arvanitakis:2021ecw}, on-shell amplitudes \cite{Bonezzi:2023xhn,Konopka:2015tta,Kunitomo:2020xrl,Nutzi:2018vkl,Arvanitakis:2019ald,Macrelli:2019afx,Jurco:2019yfd}, and correlation functions \cite{Okawa:2022sjf,Konosu:2023pal,Konosu:2023rkm,Konosu:2024dpo}, offering a more systematic and algebraic perspective on calculations traditionally carried out in quantum field theory.
Furthermore, attempts have been made to extend the analysis beyond perturbation theory \cite{Konosu:2024zrq}.
Homotopy algebras also provide a framework for demonstrating equivalences between different theories \cite{Erler:2020beb,Ando:2024pmr,JalaliFarahani:2023sfq}.

The main motivation of this paper is to develop a framework for understanding boundary string field theory (BSFT) \cite{Witten:1992qy,Witten:1992cr,Shatashvili:1993kk,Shatashvili:1993ps,Li:1993za}, which is one of the formulations of open string field theory, from the viewpoint of $A_\infty$ algebras.
BSFT is particularly well-suited for studying tachyon condensation \cite{Gerasimov:2000zp,Kutasov:2000qp,Kutasov:2000aq} and the structure of D-branes \cite{Takayanagi:2000rz, Hashimoto:2015iha}.
Unlike conventional SFT, BSFT constructs the action directly from worldsheet path integrals, resulting in a manifestly background-independent formulation.
Although BSFT has provided significant insights into the non-perturbative dynamics of open strings, its precise connection to the algebraic structures of conventional SFT remains an open question.

In this work, we propose a systematic method to construct an action that closely resembles the BSFT action by employing field redefinitions based on $A_\infty$ algebraic structures.
Specifically, we introduce a cohomomorphism that transforms the Maurer–Cartan action of a given $A_\infty$ algebra to an expression with a structure similar to that of BSFT.
We refer to this action as \textit{BSFT-like action}.
If the original Maurer–Cartan action is appropriately chosen, we expect that the BSFT-like action reproduces the BSFT action itself under a natural regularization.

The remainder of this paper is organized as follows.
In section \ref{sec:Ainfty},  we review the basics of (quantum) $A_\infty$ algebras to provide the necessary background for our construction.
Section \ref{sec:LSZ} revisits correlation functions  \cite{Okawa:2022sjf,Konosu:2023pal,Konosu:2023rkm,Konosu:2024dpo} and the LSZ reduction formula in the context of $A_\infty$ algebras \cite{Konosu:2025bnz}, with particular emphasis on off-shell amplitudes.
In Section \ref{sec:BSFT-like}, we introduce the field redefinition via the cohomomorphism and construct the BSFT-like action.  
We then investigate its properties and compare it to the standard BSFT action.
Finally, Section \ref{sec:conclusion} summarizes our results and discusses possible directions for future research.

\section{(Quantum) $A_\infty$ algebras}\label{sec:Ainfty}
In this section, as preparation for the following discussion, we review the definitions and fundamental properties of (quantum) $A_\infty$ algebras.

A (quantum) $A_\infty$ algebra is defined on a graded vector space.
We denote it as $\mathcal{H}$.
Its main components are degree-odd multi-linear products $M_{g,n}:\mathcal{H}^{\otimes n}\rightarrow \mathcal{H}$ and non-degenerate degree-odd symplectic form $\omega:\mathcal{H}^{\otimes 2}\rightarrow \mathbb{C}$, which satisfies the graded anti-symmetric property:
\begin{align}
    \omega(A_1,A_2)=-(-1)^{\deg(A_1)\deg(A_2)}\omega(A_2,A_1)\,,\label{eq:anti-sym omega}
\end{align}
where $\deg(A)$ denotes the degree of $A$.
It is convenient to introduce the bra notation:
\begin{align}
    \bra{\omega}A_1\otimes A_2=\omega(A_1,A_2)\,.
\end{align}
Let $\{e_a\}$ be a basis of $\mathcal{H}$.
For each basis vector $e_a$, there exists a dual vector $\widetilde{e}^a$ satisfying
\begin{align}
    \omega(e_a,\widetilde{e}^b)&=(-1)^{\deg(e_a)}\delta_a^b\,.\label{eq:normalization for basis}
\end{align}
The degree of $\widetilde{e}^a$ is opposite of that of $e_a$, given by
\begin{align}
    \deg(\widetilde{e}^a)=\deg(e_a)+1\quad\mod{2}\,.
\end{align}
The basis vectors $e_a$ and their duals $\widetilde{e}^a$ satisfy the following relations:
\begin{align}
    &\sum_{a}e_a\otimes\widetilde{e}^a=\sum_{a}\widetilde{e}^a\otimes e_a\,,\\
    &\sum_{a}(-1)^{\deg(\widetilde{e}^a)}e_a\otimes\bra{\omega}(\widetilde{e}^a\otimes \mathbb{I})=\sum_{a}(-1)^{\deg(e_a)}\widetilde{e}^a\otimes\bra{\omega}(e_a\otimes\mathbb{I})=\mathbb{I}\,,\label{eq:complete}
\end{align}
where $\mathbb{I}$ is denoted as the identity operator on $\mathcal{H}$.
To describe the (quantum) $A_\infty$ algebras, it is useful to introduce the tensor algebra $\mathcal{TH}$, defined as
\begin{align}
    \mathcal{TH}:=\mathcal{H}^{\otimes 0}\oplus\mathcal{H}\oplus\mathcal{H}^{\otimes 2}\oplus \mathcal{H}^{\otimes 3}\oplus\cdots\,.
\end{align}
Here, $\mathcal{H}^{\otimes 0}$ is isomorphic to $\mathbb{C}$, with a single basis vector denoted as $\bm{1}$.
The identity operator on $\mathcal{TH}$ is denoted as $\bm{I}$, and the projector from $\mathcal{TH}$ onto $\mathcal{H}^{\otimes n}$ is denoted as $\pi_n$.
Define the product
$\bigtriangledown:\mathcal{TH}\otimes'\mathcal{TH}\rightarrow\mathcal{TH}$ by
\begin{align}
    \bigtriangledown(A_1\otimes\cdots\otimes A_k\otimes'B_1\otimes\cdots\otimes B_l):=A_1\otimes\cdots\otimes A_k\otimes B_1\otimes\cdots\otimes B_l\,.
\end{align}

 On the other hand, the coproduct $\triangle:\mathcal{TH}\rightarrow \mathcal{TH}\otimes'\mathcal{TH}$ is defined by
\begin{align}
    \triangle (A_1\otimes\cdots\otimes A_n):=&\bm{1}\otimes'(A_1\otimes\cdots\otimes A_n)\nonumber\\
    &+\sum_{k=1}^{n-1}(A_1\otimes\cdots\otimes A_k)\otimes'(A_{k+1}\otimes\cdots\otimes A_n)\nonumber\\
    &+(A_1\otimes\cdots\otimes A_n)\otimes'\bm{1}\,.
\end{align}
These structures satisfy the relation:
\begin{align}
    \pi_{n+1}=\bigtriangledown(\pi_n\otimes'\pi_1)\triangle\,.\label{eq:pi_n identity}
\end{align}
The operator $\bm{M_{g,n}}:\mathcal{TH}\rightarrow\mathcal{TH}$ is defined as 
\begin{align}
    &\bm{M_{g,n}} A_1\otimes\cdots\otimes A_m\nonumber\\
    &:=
   \begin{cases}
        0&\text{if $n>m$}\,,\\
        M_{g,n}(A_1,\dots,A_n)&\text{if $n=m$}\,,\\
        \sum_{k=0}^{m-n}(-1)^{\deg(A_1)+\cdots\deg(A_k)}A_1\otimes\cdots\otimes A_k\otimes M_{g,n}(A_{k+1},\dots,A_{k+n})\\
        \qquad\qquad\qquad\qquad\qquad\quad\otimes A_{k+n+1}\otimes\cdots\otimes A_m&\text{if $n<m$}\,.
    \end{cases}\label{eq:def of coder M}
\end{align}
For the coproduct $\triangle$, the operator $\bm{M_{g,n}}$ satisfies the co-Leibniz rule:
\begin{align}
    \triangle\bm{M_{g,n}}=(\bm{I}\otimes'\bm{M_{g,n}}+\bm{M_{g,n}}\otimes'\bm{I})\triangle\,.
\end{align}
An operator satisfying the co-Leibniz rule is called a coderivation.
Given a basis vector $e_a$, the coderivation $\bm{e_a}$ is defined by
\begin{align}
    \bm{e_a}\bm{1}&:=e_a\,,\nonumber\\
    \bm{e_a}(A_1\otimes\cdots \otimes A_n)&:=\sum_{k=0}^{n}(-1)^{\deg(e_a)(\deg(A_1)+\cdots+\deg(A_k))}A_1\otimes\cdots\otimes A_k\otimes e_a\otimes A_{k+1}\otimes \cdots\otimes A_n\,,\label{eq:coder e}
\end{align}
and similarly for $\bm{\widetilde{e}^a}$.
If the operators 
\begin{align}
    \bm{M}&:=\sum_{n=0}^\infty\sum_{g=0}^\infty \hbar^{g}\bm{M_{g,n}}\,,\\
    \bm{U}&:=\frac{1}{2}\sum_{a}\bm{e_a}\bm{\widetilde{e}^a}\,,\label{eq:def of U}
\end{align}
where $\bm{M_{0,0}}=0$, satisfy the quantum $A_\infty$ relations:
\begin{align}
    (\bm{M}+i\hbar\bm{U})^2=0\,,\label{eq:quantum A relations}
\end{align}
then the algebra $(\mathcal{H},\{M_{g,n}\},\omega)$ is called a quantum $A_\infty$ algebra.
Here, $\hbar$ is a parameter, and when $\hbar=0$, these reduce to the $A_\infty$ relations:
\begin{align}
    \bm{M_0}^{\ 2}=0\,,\label{eq:A infty relations}
\end{align}
where $\bm{M_0}$ represents the $\hbar^0$ component of $\bm{M}$:
\begin{align}
    \bm{M_0}:=\sum_{n=1}^\infty\bm{M_{0,n}}\,.
\end{align}
An algebra satisfying the $A_\infty$ relations is called an $A_\infty$ algebra.
Clearly, when $\{M_{g,n}\}$ is a quantum $A_\infty$ algebra, $\{M_{0,n}\}$ is an $A_\infty$ algebra.
Additionally, if $\{M_{g,n}\}$ satisfy the cyclic property:
\begin{align}
    &\omega\big(A_1,M_{g,n}(A_2,\dots,A_{n+1})\big)\nonumber\\
    &=(-1)^{\deg(A_1)+\deg(A_{n+1})(\deg(A_1)+\cdots\deg(A_n)+1)}\omega\big(A_{n+1},M_{g,n}(A_1,\dots,A_n)\big)\,.\label{eq:cyclicity}
\end{align}
Then the algebra is called cyclic, and in this paper, we assume cyclicity implicitly.

We denote $M_{0,1}$ as $Q$ and decompose $\bm{M}$ into the coderivation $\bm{Q}$ and the remaining part $\bm{m}$:
\begin{align}
    \bm{M}=\bm{Q}+\bm{m}\,.
\end{align}
The operator $Q$ is nilpotent:
\begin{align}
    Q^2=0\,.
\end{align}
Let us consider the degree-odd homotopy operator $h$ and a projector $P$ satisfying the following conditions:
\begin{align}
    Qh+hQ=\mathbb{I}-P\,,\quad hP=0\,,\quad Ph=0\,,\quad h^2=0\,,\quad QP=0\,,\quad PQ=0\,.\label{eq:sdl}
\end{align}
The last two conditions imply that the projection $P$ is onto the $Q$-cohomology.
In this case, $Qh$ and $hQ$ are also projections, so the space $\mathcal{H}$ can be decomposed into three subspaces: 
\begin{align}
    \mathcal{H}^{(P)}:=P\mathcal{H}\,,\quad\mathcal{H}^{(Qh)}:=Qh\mathcal{H}\,,\quad\mathcal{H}^{(hQ)}:=hQ\mathcal{H}\,.
\end{align}
The subspace $\mathcal{H}^{(P)}$ represents the $Q$-cohomology, which corresponds to the space of on-shell degrees of freedom.
Additionally, in this paper, we assume that $h$ satisfies the following property,
\begin{align}
    \omega(A_1,hA_2)=(-1)^{\deg(A_1)}\omega(hA_1,A_2)\,.\label{eq:h property}
\end{align}
In this setting, for $e_a \in \mathcal{H}^{(P)}$, its dual lies in $\mathcal{H}^{(P)}$. In contrast, for $e_a \in \mathcal{H}^{(Qh)}$, its dual lies in $\mathcal{H}^{(hQ)}$, and vice versa.
Specifically, this can be verified as follows.
For a given $e_a$, its dual $\widetilde{e}^a$ is defined by the condition $\omega(e_a, \widetilde{e}^a)=1$.
Let us consider the case where $e_a \in \mathcal{H}^{(P)}$.
If $\widetilde{e}^a \in \mathcal{H}^{(Qh)}$, we can write $\omega(e_a, \widetilde{e}^a) = \omega(e_a, Qh\widetilde{e}^a)$.
However, due to the cyclicity of $Q$ \eqref{eq:cyclicity}, we can make $Q$ act on $e_a$, and thus the term vanishes.
Similarly, if $\widetilde{e}^a \in \mathcal{H}^{(hQ)}$, we have $\omega(e_a, \widetilde{e}^a) = \omega(e_a, hQ\widetilde{e}^a)$.
By the property \eqref{eq:h property}, $h$ acts on $e_a$, which also makes the term vanish.
Therefore, the dual $\widetilde{e}^a$ of an element $e_a \in \mathcal{H}^{(P)}$ must belong to $\mathcal{H}^{(P)}$.
The cases for $e_a \in \mathcal{H}^{(Qh)}$ and $e_a \in \mathcal{H}^{(hQ)}$ can be verified in a similar manner.

Operators $h$ and $P$ can be extended to operators $\bm{h}$ and $\bm{P}$ on $\mathcal{TH}$, respectively, as follows:
\begin{align}
    \bm{h}&:=\sum_{k=0}^\infty \sum_{l=0}^\infty\big(\mathbb{I}^{\otimes k}\otimes h\otimes P^{\otimes l}\big)\pi_{k+l+1}\,,\label{eq:def of h}\\
    \bm{P}&:=\sum_{k=0}^\infty P^{\otimes k}\pi_k\,.
\end{align}
These operators satisfy the relations:
\begin{align}
    \bm{hQ}+\bm{Qh}=\bm{I}-\bm{P}\,,\quad \bm{hP}=0\,,\quad \bm{Ph}=0\,,\quad \bm{h}^2=0\,,\quad \bm{QP}=0\,,\quad \bm{PQ}=0\,.
\end{align}

Here for any two maps $\bm{A}$ and $\bm{B}$ on $\mathcal{TH}$, we define a map $\bm{A}\otimes \bm{B}:\mathcal{TH}\rightarrow\mathcal{TH}$ by\footnote{Be careful to distinguish the tensor product $\otimes$ on $\mathcal{H}$ from the tensor product $\otimes'$ on $\mathcal{TH}$.
The map $\bm{A}\otimes\bm{B}$ is not a map acting on $\mathcal{TH}\otimes'\mathcal{TH}$; rather, it acts on $``\mathcal{TH}\otimes\mathcal{TH}''=\mathcal{TH}$.
(The quotation marks indicate that writing $\mathcal{TH}\otimes\mathcal{TH}$ using the tensor product on $\otimes$ is not strictly correct.)
More precisely, we first project $\mathcal{TH}$ onto $\mathcal{H}^{\otimes n}$, take the tensor product using $\otimes$ there.}
\begin{align}
    \bm{A}\otimes \bm{B}:=\bigtriangledown(\bm{A}\otimes'\bm{B})\triangle\,.\label{A otimes B}
\end{align}
Moreover, using that the product $\bigtriangledown$ and the coproduct $\triangle$ satisfy
\begin{align}
    \bigtriangledown(\bigtriangledown\otimes'\bm{I})&=\bigtriangledown(\bm{I}\otimes'\bigtriangledown)\,,\\
    (\triangle\otimes'\bm{I})\triangle&=(\bm{I}\otimes'\triangle)\triangle\,,
\end{align}
one finds that for three maps $\bm{A}, \bm{B}$ and $\bm{C}$ on $\mathcal{TH}$,
\begin{align}
    (\bm{A}\otimes\bm{B})\otimes\bm{C}=\bm{A}\otimes(\bm{B}\otimes\bm{C})\,.
\end{align}

The Maurer-Cartan action is given by
\begin{align}
    S := -\sum_{n=0}^\infty \sum_{g=0}^\infty \frac{\hbar^{g}}{n+1}\omega\big(\Phi,M_{g,n}(\underbrace{\Phi,\dots,\Phi}_n)\big)\,.\label{eq:action}
\end{align}
The field $\Phi$ is a degree-even element of a half-dimensional subspace $\mathcal{H}_{\mathrm{g.f.}}\subset\mathcal{H}$.
As for the choice of $\mathcal{H}_{\mathrm{g.f.}}$, we proceed as follows.
On the subspace of $\mathcal{H}$ excluding $\mathcal{H}^{(P)}$, we impose the gauge fixing condition $h\Phi=0$ and select $\mathcal{H}^{(hQ)}$.
On $\mathcal{H}^{(P)}$, we impose some additional gauge fixing condition by hand to pick a half-dimensional subspace $\mathcal{H}^{(P)}_{\mathrm{g.f.}}$.
Importantly, we choose the gauge-fixing condition so that, for any basis elements $e_a$, the dual basis element $\widetilde{e}^a$ lies in the complementary subspace $\mathcal{H}^{(P)}\setminus\mathcal{H}^{(P)}_{\mathrm{g.f.}}$.
In summary, $\mathcal{H}_{\mathrm{g.f.}}$ can be decomposed as
\begin{align}
    \mathcal{H}_{\mathrm{g.f.}}=\mathcal{H}^{(P)}_{\mathrm{g.f.}}\oplus \mathcal{H}^{(hQ)}\,,
\end{align}
and $\Phi$ can be expanded in the basis as 
\begin{align}
    \Phi = \sum_{e_a\in \mathcal{H}_{\mathrm{g.f.}}}\phi^a e_a\,.
\end{align}
Since $h$ annihilates both $\mathcal{H}^{(P)}_{\mathrm{g.f.}}\subset\mathcal{H}^{(P)}$ and $\mathcal{H}^{(hQ)}$, it follows that the condition
\begin{align}
    h\Phi=0\,,
\end{align}
holds as a whole.
Note, however, that satisfying $h\Phi=0$ does not necessarily imply that $\mathcal{H}_{\mathrm{g.f.}}$.

\section{Correlation functions and LSZ reduction formula}\label{sec:LSZ}

In this section, we review correlation functions \cite{Okawa:2022sjf,Konosu:2023pal,Konosu:2023rkm,Konosu:2024dpo} and the LSZ reduction formula \cite{Konosu:2025bnz} in the framework of (quantum) $A_\infty$ algebras as a foundation for the following discussion.
However, the previous works focused on the case where $P=0$, while we do not impose this restriction here.

\subsection{Correlation functions}

In \cite{Okawa:2022sjf,Konosu:2023pal,Konosu:2023rkm,Konosu:2024dpo}, 
correlation functions were considered in the case where $P=0$.
In this case, the space $\mathcal{H}$ is decomposed into two subspaces $\mathcal{H}^{(hQ)}$ and $\mathcal{H}^{(Qh)}$,  and the projection $\bm{P}$ is given by 
\begin{align}
    \bm{P}=\pi_0\,.
\end{align} 
The correlation functions for the theory defined by the action \eqref{eq:action} are given by
\begin{align}
    \langle\phi^{a_1}\cdots\phi^{a_n}\rangle=(-1)^n\bra{\omega}(\widetilde{e}^{a_1}\otimes\pi_1)\otimes\cdots\otimes\bra{\omega}(\widetilde{e}^{a_n}\otimes\pi_1)\pi_n\frac{1}{\bm{I}+\bm{h}(\bm{m}+i\hbar\,\bm{U})}\bm{1}\,,\label{eq:correlation function}
\end{align}
for $\widetilde{e}^{a_i}\in\mathcal{H}^{(Qh)}$.
Here, the operator $\frac{1}{\bm{I}+\bm{h}(\bm{m}+i\hbar\,\bm{U})}$ is defined by the formal power series
\begin{align}
    \frac{1}{\bm{I}+\bm{h}(\bm{m}+i\hbar\,\bm{U})}:=\bm{I}+\sum_{n=1}^\infty (-1)^n\big(\bm{h}(\bm{m}+i\hbar\,\bm{U})\big)^n\,.
\end{align}
For notational simplicity, we denote this operator as $\bm{f}$.
The formal expansion of $\bm{f1}$ corresponds to the Feynman diagram expansion. 
We now comment on the operator $\bra{\omega}(\widetilde{e}^a\otimes \pi_1)$.
It acts as a map $\mathcal{H} \to \mathbb{C}$.
The tensor product of such maps, $\bra{\omega}(\widetilde{e}^a\otimes \pi_1)\otimes\bra{\omega}(\widetilde{e}^b\otimes \pi_1)$, maps $\mathcal{H}\otimes\mathcal{H} \to \mathbb{C}$ and is defined for inputs $A, B \in \mathcal{H}$ by
\begin{align}
    \bra{\omega}(\widetilde{e}^a\otimes \pi_1)\otimes\bra{\omega}(\widetilde{e}^b\otimes \pi_1) A\otimes B=(-1)^{\deg(A)(\deg(\widetilde{e}^b)+1)}\bra{\omega}(\widetilde{e}^a\otimes A)\otimes\bra{\omega}(\widetilde{e}^b\otimes B)\,.
\end{align}
Here, the sign factor on the right-hand side originates from the inputs moving past $\bra{\omega}$ and $\widetilde{e}^b$.
We recall that $\bra{\omega}$ is degree-odd.
The operator $\bm{f}$ satisfies the following relations: \footnote{The relation between the correlators characterized by this Schwinger--Dyson equation and the correlation functions based on the factorization algebra of Costello and Gwilliam \cite{Costello:2016vjw, Costello:2021jvx} was discussed in Appendix B of \cite{Konosu:2023pal}.
We should note, however, that the correlation functions defined in this way are not unique. There is a freedom in the choice of the homotopy operator $h$, which corresponds to the choice of Green's function.}
\begin{align}
    \big(\bm{M}+i\hbar\,\bm{U}\big)\bm{f}\bm{1}=0\,.\label{eq:algebraicSD}
\end{align}
By acting with $\bra{\omega}(\widetilde{e}^{a_1}\otimes\pi_1)\otimes\cdots\otimes\bra{\omega}(\widetilde{e}^{a_n}\otimes\pi_1)\otimes\bra{\omega}(e_b\otimes\pi_1)$ for $\widetilde{e}^{a_i}\in \mathcal{H}^{(Qh)}, e_b\in\mathcal{H}^{(hQ)}$ on this identity, one can verify that this is equivalent to the Schwinger-Dyson equation \cite{Konosu:2024dpo}.

In this paper, however, we consider the case $P\neq 0$.
In this case, \eqref{eq:algebraicSD} is modified as
\begin{align}
    \big(\bm{M}+i\hbar\,\bm{U}\big)\bm{f}\bm{1}=\bm{f}\bm{P}(\bm{m}+i\hbar\,\bm{U})\bm{f1}\,.\label{ASD Pneq0}
\end{align}
Due to the right-hand side, $\bm{f}\bm{1}$ does not satisfy the Schwinger-Dyson equation in the usual sense.
Therefore, when $P\neq0$, $\bm{f}\bm{1}$ should not be regarded as defining correlation functions in the strict sense.
Nevertheless, this does not cause any difficulty for the arguments below.
The formal expansion of $\bm{f}\bm{1}$ still generates perturbative Feynman diagrams in the same way as in the case $P=0$.
Thus, the extraction of the connected part from $\bm{f}\bm{1}$ and the subsequent LSZ reduction can be carried out almost in parallel with the analysis of \cite{Konosu:2025bnz}, where the discussion starts from $\bm{f}\bm{1}$ with $P=0$.
For brevity, we refer to the quantities defined by \eqref{eq:correlation function} as correlation functions also when $P\neq0$, by abuse of terminology.
This terminology is only a convention within the present paper and should not be understood as assigning them the strict interpretation of conventional correlation functions.

\subsection{Connected correlation functions}\label{subsec:connected correlation functions}
To identify the part of $\bm{f1}$ that corresponds to connected correlation functions, let us analyze the diagrams generated by $\bm{f1}$.
For simplicity, we consider the case where $\bm{m}=\bm{M_{0,2}}$.
For example, let us calculate the two-point correlation function up to order $\mathcal{O}(\hbar^2)$.
First, the formal expansion of the operator $\bm{f}$ is given by
\begin{align}
    \bm{f}=&\bm{I}-\bm{h}(\bm{M_{0,2}}+i\hbar\,\bm{U})+\Big(\bm{h}(\bm{M_{0,2}}+i\hbar\,\bm{U})\Big)^2\nonumber\\
    &-\Big(\bm{h}(\bm{M_{0,2}}+i\hbar\,\bm{U})\Big)^3+\Big(\bm{h}(\bm{M_{0,2}}+i\hbar\,\bm{U})\Big)^4+\cdots\,.
\end{align}
Among these terms, the contribution that survives up to order $\mathcal{O}(\hbar^2)$, when acted upon by $\pi_2$ from the left and $\bm{1}$ from the right, is reduced to
\begin{align}
    \pi_2\bm{f1}=\pi_2\Big(-i\hbar\,\bm{hU1}+(i\hbar)^2\,\big(\bm{hM_{0,2}hM_{0,2}hUhU1}+\bm{hM_{0,2}hUhM_{0,2}hU1}\big)\Big)+\mathcal{O}(\hbar^3)\,.
\end{align}
We now examine the first term, $\bm{h}\bm{U}\bm{1}$.
From the definitions of $\bm{h}$ \eqref{eq:def of h} and $\bm{U}$ \eqref{eq:def of U}, this term can be written as
\begin{align}
    \bm{hU1}=\sum_{a}\Big(he_a\otimes P\widetilde{e}^a+e_a\otimes h\widetilde{e}^a\Big)\,.
\end{align}
However, the sub-term $h e_a \otimes P \widetilde{e}^a$ vanishes.
This is because when $e_a$ is not annihilated by $h$ (i.e., $e_a \in \mathcal{H}^{(Qh)}$), its dual $\widetilde{e}^a$ belongs to the space $\mathcal{H}^{(hQ)}$, and is therefore annihilated by $P$.
Note that $e_a \otimes h\widetilde{e}^a$ is an element of $\mathcal{H}^{(hQ)} \otimes \mathcal{H}^{(hQ)}$, and both tensor factors are annihilated by $h$ and $P$. Calculating the remaining terms, we obtain the following result:
\begin{align}
    \pi_2\bm{f1}=&-i\hbar\sum_a e_a\otimes h\widetilde{e}^a \nonumber\\
    &+(i\hbar)^2\sum_{a,b}\Big[e_a\otimes hM_{0,2}\big(h\widetilde{e}^a, hM_{0,2}(e_b,h\widetilde{e}^b)\big)\nonumber\\
    &\qquad\qquad+(-1)^{\deg(e_a)\deg(e_b)}e_a\otimes hM_{0,2}\big(e_b,hM_{0,2}(h\widetilde{e}^a,h\widetilde{e}^b)\big)\nonumber\\
    &\qquad\qquad+e_a\otimes hM_{0,2}\big(e_b,hM_{0,2}(h\widetilde{e}^b,h\widetilde{e}^a)\big)\nonumber\\
    &\qquad\qquad+e_a\otimes hM_{0,2}\big(hM_{0,2}(e_b,h\widetilde{e}^b),h\widetilde{e}^a\big)\nonumber\\
    &\qquad\qquad+hM_{0,2}(e_a,h\widetilde{e}^a)\otimes hM_{0,2}(e_b,h\widetilde{e}^b)
    \Big]\nonumber\\
    &+\mathcal{O}(\hbar^3)\,.\label{eq:two point correlations}
\end{align}
While this result coincides with that obtained for $P=0$, we emphasize that the calculation presented here was performed for the case $P \neq 0$.
Among these terms, the following term represents the disconnected diagrams:
\begin{align}
    (i\hbar)^2\sum_{a,b}hM_{0,2}(e_a,h\widetilde{e}^a)\otimes hM_{0,2}(e_b,h\widetilde{e}^b)\,,
\end{align}
while the remaining terms correspond to the connected parts.
A feature of the connected parts is that the operation $hM_{0,2}$ appears only in the rightmost position.
This property holds in the general case.

To make this feature of the connected part more transparent, we calculate $\pi_{n+1}\bm{f1}=\bigtriangledown(\pi_n\otimes'\pi_1)\triangle \bm{f1}$. 
The action of the coproduct $\triangle$ on $\bm{hm}$ and $\bm{hU}$ is given by:
\begin{align}
    \triangle\bm{hm}=&(\bm{h}\bm{m}\otimes'\bm{P}-\bm{m}\otimes'\bm{h}+\bm{I}\otimes'\bm{hm})\triangle\,,\label{eq:triangle action hm}\\
    \triangle\bm{hU}=&(\bm{hU}\otimes'\bm{P}-\bm{U}\otimes'\bm{h}+\bm{I}\otimes'\bm{hU}\nonumber\\
    &+\sum_a(-1)^{\deg(e_a)}\bm{he_a}\otimes' \bm{P\widetilde{e}^a}+\sum_a(-1)^{\deg(e_a)}\bm{e_a}\otimes' \bm{h\widetilde{e}^a})\triangle\,.\label{eq:triangle action hU}
\end{align}
When calculating $\triangle\bm{f}\bm{1}$, since these operators act on the spaces annihilated by $h$, the second term in \eqref{eq:triangle action hm}, and both the second term and fourth term in \eqref{eq:triangle action hU}, vanish.
As a result, we can express $\triangle\bm{f1}$ as:
\begin{align}
    \triangle\bm{f1}=
    &\sum_{n=0}^\infty(-1)^n\Big[\bm{h}(\bm{m}+i\hbar\,\bm{U})\otimes'\bm{P}+\bm{I}\otimes'\bm{h}(\bm{m}+i\hbar\,\bm{U})\nonumber\\
    &\qquad\qquad+i\hbar\sum_a(-1)^{\deg(e_a)}\bm{e_a}\otimes'\bm{h}\bm{\widetilde{e}^a}\Big]^n\,(\bm{1}\otimes'\bm{1})\,.
\end{align}
For brevity, let us define:
\begin{subequations}
    \begin{align}
    A&:=\bm{h}(\bm{m}+i\hbar\,\bm{U})\otimes'\bm{P}\,,\\
    B&:=\bm{I}\otimes'\bm{h}(\bm{m}+i\hbar\,\bm{U})\,,\\
    C&:=i\hbar\sum_a(-1)^{\deg(e_a)}\bm{e_a}\otimes'\bm{h}\bm{\widetilde{e}^a}\,.
\end{align}
\end{subequations}
Since $AB=AC=0$, the expression simplifies to:
\begin{align}
    \triangle\bm{f1}&=\sum_{n=0}^\infty \big(-A-B-C\big)^n\bm{1}\otimes'\bm{1}\,,\nonumber\\
    &=\sum_{k=0}^\infty\big(-B-C\big)^k\sum_{l=0}^\infty(-A)^l\bm{1}\otimes'\bm{1}\,,\nonumber\\
    &=\frac{1}{1+B+C}\frac{1}{1+A}\bm{1}\otimes'\bm{1}\,.
\end{align}
Furthermore, we can expand:
\begin{align}
    \frac{1}{1+B+C}=\frac{1}{1+\frac{1}{1+B}C}\frac{1}{1+B}=\sum_{k=0}^\infty \big[-\frac{1}{1+B}C\big]^k\frac{1}{1+B}\,.
\end{align}
Using $\frac{1}{1+A}=\bm{f}\otimes'\bm{P}$ and $\frac{1}{1+B}=\bm{I}\otimes'\bm{f}$\,, we obtain:
\begin{align}
     \triangle\bm{f1}=\sum_{k=0}^\infty\sum_{a_1,\dots,a_k}(-1)^{k+\deg(e_{a_1})+\cdots+\deg(e_{a_k})}(i\hbar)^k\bm{e_{a_1}}\cdots\bm{e_{a_k}}\bm{f1}\otimes'\bm{f}\bm{h\widetilde{e}^{a_k}}\cdots\bm{f}\bm{h\widetilde{e}^{a_1}}\bm{f1}\,.
\end{align}
Thus, the expression for $\pi_{n+1}\bm{f1}$ becomes:
\begin{align}
    \pi_{n+1}\bm{f1}=\bigtriangledown(\pi_n\otimes'\pi_1)\triangle\bm{f1}=\sum_{k=0}^\infty\sum_{a_1,\dots,a_k}&(-1)^{k+\deg(e_{a_1})+\cdots+\deg(e_{a_k})}(i\hbar)^k\pi_n\bm{e_{a_1}}\cdots\bm{e_{a_k}}\bm{f1}\nonumber\\
     &\otimes\pi_1\bm{f}\bm{h\widetilde{e}^{a_k}}\cdots\bm{f}\bm{h\widetilde{e}^{a_1}}\bm{f1}\,.
\end{align}
From this expression, it is clear that the connected part, characterized by the absence of $\bm{hm}$ except in the rightmost position, is given by:
\begin{align}
    \sum_{n=0}^\infty \sum_{a_1,\dots,a_n}(-1)^{n+\deg(e_{a_1})+\cdots+\deg(e_{a_n})}(i\hbar)^n\bm{e_{a_1}}\cdots\bm{e_{a_n}}\bm{1}\otimes\pi_1\bm{f}\bm{h\widetilde{e}^{a_n}}\cdots\bm{f}\bm{h\widetilde{e}^{a_1}}\bm{f1}\,.\label{eq:connected part}
\end{align}
The connected part of the $(n+1)$-point correlation function can then be obtained by acting with $(-1)^{n+1}\bra{\omega}(\widetilde{e}^{a_1}\otimes\pi_1)\otimes\cdots\otimes\bra{\omega}(\widetilde{e}^{a_{n+1}}\otimes\pi_1)$ on this expression:
\begin{align}
    &\langle\phi^{a_1}\cdots\phi^{a_{n+1}}\rangle_{\mathrm{conn}}\nonumber\\
    &=(i\hbar)^{n}\sum_{\sigma\in S_{n}}(-1)^{n+1+\deg(e_{a_{n+1}})\big(\deg(e_{a_1})+\cdots+\deg(e_{a_{n}})\big)+\epsilon(\sigma)}\omega\Big(\widetilde{e}^{a_{n+1}},\pi_1\bm{f}\bm{h\widetilde{e}^{a_{\sigma(1)}}}\cdots\bm{f}\bm{h\widetilde{e}^{a_{\sigma(n)}}}\bm{f1}\Big)\,,\label{eq:connected correlation functions}
\end{align}
where $S_{n}$ is the symmetric group of degree $n$, and $(-1)^{\epsilon(\sigma)}$ is the sign from permuting $e_{a_1},\dots,e_{a_{n}}$ into the order $e_{a_{\sigma(1)}},\dots,e_{a_{\sigma(n)}}$.
We note that \eqref{eq:normalization for basis} was used in the derivation of this expression.
This expression includes the connected correlation functions for all cyclic permutations of $\phi^{a_1},\dots,\phi^{a_{n+1}}$.
In particular, the component corresponding to the ordering $\phi_1,\dots,\phi_{n+1}$ along the disk boundary (clockwise) is given by:
\begin{align}
    &\langle\langle\phi^{a_1}\cdots\phi^{a_{n+1}}\rangle\rangle_{\mathrm{conn}}
    :=(i\hbar)^{n}(-1)^{n+1+\rev{e_{a_1},\dots,e_{a_{n+1}}}}
    \omega\Big(\widetilde{e}^{a_{n+1}},\pi_1\bm{f}\bm{h\widetilde{e}^{a_n}}\cdots\bm{f}\bm{h\widetilde{e}^{a_1}}\bm{f1}\Big)\,,\label{eq:orderd connected correlation functions}
\end{align}
where $(-1)^{\rev{e_{a_1},\dots e_{a_{n+1}}}}$ is the sign arising from reversing the order of  $e_{a_1},\dots,e_{a_{n+1}}$.
In the next subsection, we amputate the external lines according to the LSZ reduction formula to compute the off-shell amplitude. 

\subsection{Tree-level off-shell amplitude}

Let us now consider the tree-level off-shell amplitude.
At the tree-level, the connected correlation function reduces to
\begin{align}
     \langle\langle\phi^{a_1}\cdots\phi^{a_{n+1}}\rangle\rangle_{\mathrm{conn}}^{\mathrm{tree}}=(i\hbar)^{n}(-1)^{n+1+\rev{e_{a_1},\dots,e_{a_{n+1}}}}\omega\Big(\widetilde{e}^{a_{n+1}},\pi_1\bm{f_0}\bm{h\widetilde{e}^{a_{n}}}\cdots\bm{f_0}\bm{h\widetilde{e}^{a_{1}}}\bm{1}\Big)\,,\label{eq:tree level connected correlation function}
 \end{align}
 where $\bm{f_0}$ represents the $\mathcal{O}(\hbar^0)$ component of $\bm{f}$, defined as
 \begin{align}
     \bm{f_0}&:=\frac{1}{\bm{I}+\bm{h}\bm{m_0}}\,,
\end{align}
with
\begin{align}
     \bm{m_0}&:=\sum_{n=2}^\infty \bm{M_{0,n}}=\bm{M_0}-\bm{Q}\,,
 \end{align}
 and we used $\bm{f_01}=\bm{1}$.
It is evident that the homotopy operator $h$ acts on every $\widetilde{e}^{a_i}$.
The homotopy operator $h$ corresponds to the propagator and is precisely the external line at the tree-level.
To obtain the off-shell amplitude, we amputate these external lines.

First, consider the simple case where $P=0$.
In this case, $\Phi$ satisfying $h\Phi=0$ belongs to $\mathcal{H}^{(hQ)}$.
Thus, by replacing $\widetilde{e}^{a_i}$ with $Q\Phi_i$, the external propagator $h$ can be amputated.
By choosing the overall sign and $i\hbar$ factor to match those of the minimal model for later comparison, the off-shell amplitude is written as:
\begin{align}
-\omega\Big(Q\Phi_{n+1},\pi_1\bm{f_0}\bm{h}\bm{D}(Q\Phi_n)\cdots\bm{f_0}\bm{h}\bm{D}(Q\Phi_1)\bm{1}\Big)\,,\label{eq:off shell amplitude}
\end{align}
where $\bm{D}(A)$ for $A\in \mathcal{H}$ is a coderivation generated by $A$ defined by
\begin{align}
    &\bm{D}(A)\bm{1}:=A\,,\nonumber\\
    &\bm{D}(A)\big(B_1\otimes\cdots\otimes B_n\big)\nonumber\\
    &\quad:=\sum_{k=0}^n(-1)^{\deg(A)\big(\deg(B_1)+\cdots+\deg(B_n)\big)}B_1\otimes\cdots\otimes B_k\otimes A\otimes B_{k+1}\otimes \cdots\otimes B_n\,,\label{eq:def of D}
\end{align}
and $\Phi_1,\dots\Phi_{n+1}$ are degree-even elements of $\mathcal{H}^{(hQ)}$.
By using this definition,  we can also write $\bm{e}_a = \bm{D}(e_a)$.
The homotopy operator $\bm{h}$ in front of $\bm{D}(Q\Phi_i)$ contributes only when acting on $Q\Phi_i$; otherwise, it vanishes. 
Since $hQ\Phi_i=\Phi_i$, which follows from the relation $Qh+hQ=\mathbb{I}-P$, the condition $h\Phi_i=0$ and the fact that we are currently considering $P=0$, the external propagators are effectively amputated.

In the general case where $P\neq0$, the relation $hQ\Phi=(\mathbb{I}-P)\Phi$ holds, meaning that $hQ\Phi$ is not necessarily equal to $\Phi$.
Therefore, some modification is needed to properly amputate the external lines.
Consider introducing $\epsilon$-dependent operators $Q_\epsilon$, satisfying
\begin{align}
    \lim_{\epsilon\rightarrow0}Q_{\epsilon}=Q,
\end{align}
where $\epsilon$ is a parameter. 
We also introduce $h_\epsilon$ with $\epsilon>0$ satisfying
\begin{align}
    Q_\epsilon h_\epsilon+h_\epsilon Q_\epsilon=\mathbb{I}\,.\label{eq:Hodge-Kodaira epsilon}
\end{align}
Additionally, we assume that $h_\epsilon$ satisfies 
\begin{align}
    h_\epsilon\Phi=0\,,\label{h_epsilonPhi=0}
\end{align} for $\Phi\in\mathcal{H}_{\mathrm{g.f.}}$.
As seen in the explicit examples below, $h_\epsilon$ does not in general connect smoothly to $h$ on $\mathcal{H}^{(P)}$, so taking the limit $\epsilon\rightarrow0$ is not a well-defined operation.
Nevertheless, in this paper we use the notation $\lim_{\epsilon\rightarrow0}h_\epsilon$ to denote the formal replacement of $h_\epsilon$ by $h$.

A concrete example can be given in the case of scalar theory, where $Q$ and $h$ are defined as follows:
\begin{align}
    Q\,e(x) = (-\partial^2+m^2)\widetilde{e}(x)\,,&\qquad Q\,\widetilde{e}(x)=0\,,\\
    h\,e(x)=0,&\qquad h\,\widetilde{e}(x)=\int \dd^dy\frac{\dd^dk}{(2\pi)^d}\frac{e^{ik(x-y)}}{k^2+m^2}e(y)\,,
\end{align}
where $e(x)\in \mathcal{H}^{(hQ)}$ and $\widetilde{e}(x)\in \mathcal{H}^{(Qh)}$.
For elements of $\mathcal{H}^{(P)}$, $Q$ and $h$ vanish by definition.
For these operators, $\epsilon$-dependent operators $Q_\epsilon$ and $h_\epsilon$ can be defined as follows:
\begin{align}
    Q_\epsilon\,e(x) = (-\partial^2+m^2-i\epsilon)\widetilde{e}(x)\,,&\qquad Q_\epsilon\,\widetilde{e}(x)=0\,,\\
    h_\epsilon\,e(x)=0,&\qquad h_\epsilon\,\widetilde{e}(x)=\int \dd^dy\frac{\dd^dk}{(2\pi)^d}\frac{e^{ik(x-y)}}{k^2+m^2-i\epsilon}e(y)\,,
\end{align}
again with $e(x)\in \mathcal{H}_{\mathrm{g.f.}}$ and $\widetilde{e}(x)\in \mathcal{H}\setminus\mathcal{H}_{\mathrm{g.f.}}$.
In particular, for $e(x)\in \mathcal{H}^{(P)}_{\mathrm{g.f.}}$ and $\widetilde{e}(x)\in\mathcal{H}\setminus\mathcal{H}^{(P)}_{\mathrm{g.f.}}$, we have
\begin{align}
    Q_\epsilon\,e(x) = -i\epsilon\,\widetilde{e}(x)\,,&\qquad Q_\epsilon\,\widetilde{e}(x)=0\,,\\
    h_\epsilon\,e(x)=0,&\qquad h_\epsilon\,\widetilde{e}(x)=\frac{1}{-i\epsilon}e(x)\,,
\end{align}
so unlike $Q$ and $h$, it acts nontrivially.
These operators are manifestly nilpotent and satisfy the required conditions.
Of course, one can construct $\epsilon$-dependent operators in string field theory.
For instance, in bosonic open string field theory with Siegel gauge, $b_0\Phi=0$, the kinetic term takes the form
\begin{align}
    \frac{1}{2}\omega\big(\Phi,c_0L_0\Phi\big)\,.
\end{align}
Accordingly, $Q$ and $h$ are defined as 
\begin{align}
    Q=c_0L_0\,,\qquad h=\frac{b_0}{L_0}\big(\mathbb{I}-P_{\ker{L_0}}\big)\,,
\end{align}
where, $P_{\ker{L_0}}$ denotes the projection onto $\ker{L_0}$.
The corresponding operators $Q_\epsilon$ and $h_\epsilon$ are then given by
\begin{align}
    Q_\epsilon=c_0(L_0-i\epsilon)\,,\qquad h_\epsilon = \frac{b_0}{L_0-i\epsilon}\,.
\end{align}

With these definitions, the tree-level off-shell amplitude is given by
\begin{align}
    -\lim_{\epsilon\rightarrow 0}\,\omega\Big(Q_\epsilon\Phi_{n+1},\pi_1\bm{f^{(\epsilon)}_0}\bm{h_\epsilon}\bm{D}(Q_\epsilon\Phi_n)\cdots\bm{f^{(\epsilon)}_0}\bm{h_\epsilon}\bm{D}(Q_\epsilon\Phi_1)\bm{1}\Big)\,,\label{eq:off-shell amplitude with epsilon}
\end{align}
where $\Phi_1,\dots,\Phi_{n+1}$ are degree-even elements of $\mathcal{H}_{\mathrm{g.f.}}$.
Note that these elements vanish under the action of $h_\epsilon$ by assumption \eqref{h_epsilonPhi=0}.
The operator $\bm{f^{(\epsilon)}_0}$ is defined by replacing $h$ with $h_\epsilon$ in the definition of $\bm{f_0}$, and $\bm{h_\epsilon}$ for $\epsilon>0$ is defined by
\begin{align}
    \bm{h_\epsilon}:=\sum_{k=0}^\infty \big(\mathbb{I}^{\otimes k}\otimes h_\epsilon\big)\pi_{k+1}\,.\label{eq:def of bmhepsilon}
\end{align}
It is crucial to note that although $h_\epsilon$ approaches $h$ in the limit $\epsilon \rightarrow 0$, the corresponding lifted operators $\bm{h}_\epsilon$ and $\bm{h}$ are distinct.
The definition of $\bm{h}_\epsilon$ is \eqref{eq:def of bmhepsilon}
, whereas the definition of $\bm{h}$ is \eqref{eq:def of h}.
These two operators are different because we are now considering the case $P\neq0$. 
This modification corresponds to the standard $i\epsilon$-prescription in quantum field theory.
In the expression \eqref{eq:off-shell amplitude with epsilon}, the operator $\bm{h}_\epsilon$ preceding $\bm{D}(Q_\epsilon\Phi_i)$ only acts on $Q_\epsilon\Phi_i$, vanishing otherwise, similar to the case $P=0$.
Using \eqref{eq:Hodge-Kodaira epsilon} and \eqref{h_epsilonPhi=0} we obtain $h_\epsilon Q_\epsilon \Phi=\Phi$, which realizes the amputation of the external lines.
Although one might initially expect the term to vanish, since $\lim_{\epsilon\rightarrow 0} Q_\epsilon \Phi_i = Q \Phi_i$ when $\Phi_i \in \mathcal{H}^{(P)}$, this issue is avoided because the limit is taken only after the cancellation between $\bm{h}_\epsilon$ and $Q_\epsilon$ has occurred.
Note that $\bm{h_\epsilon D}(Q_\epsilon\Phi_i)$ is not identical to $\bm{D}(\Phi_i)$, because they are distinct operators.
For example:
\begin{align}
    \bm{h_\epsilon}\bm{D}(Q_\epsilon\Phi_2)\bm{h_\epsilon}\bm{D}(Q_\epsilon\Phi_1)\bm{1}=\Phi_1\otimes\Phi_2\,,
\end{align}
whereas
\begin{align}
    \bm{D}(\Phi_2)\bm{D}(\Phi_1)\bm{1}=\Phi_1\otimes\Phi_2+\Phi_2\otimes\Phi_1\,.
\end{align}
The operator $\bm{h_\epsilon}\bm{D}(Q_\epsilon\Phi_i)$ can thus be interpreted as appending $\Phi_i$ to the rightmost position.

For $n=1$, we obtain
\begin{align}
    -\lim_{\epsilon\rightarrow 0}\,\omega\big(Q_\epsilon\Phi_2,h_\epsilon Q_\epsilon\Phi_1\big)=\omega\big(\Phi_2,Q\Phi_1\big)\,,
\end{align}
which can be interpreted as a ``two-point amplitude''.
Naively, since $PQ=QP=0$, this does not contribute to on-shell amplitudes.
However, in general, two-point amplitudes do not vanish.
In particular, the existence of non-vanishing on-shell two-point amplitudes in string theory has been pointed out in \cite{Erbin:2019uiz}.
This can be understood through a toy model with the following setup:
\begin{subequations}
\begin{align}
    \Phi&:=\int \dd^dk\,\phi(k)e(k)\,,\\
    Pe(k)&:=\delta^d(k^2+m^2)e(k)\,,\\
    Qe(k)&:=(k^2+m^2)e(k)\,,\\
    \omega\big(e(k),\widetilde{e}(k')\big)&:=\delta^d(k-k')\,.
\end{align}
\end{subequations}
In this setup, the two-point on-shell amplitude is computed as
\begin{align}
    \omega\big(P\Phi_2,QP\Phi_1\big)&=\int\dd^dk\dd^dk'\,\phi_2(k)\phi_1(k')\delta^d(k^2+m^2)\delta^d(k'^2+m^2)(k'^2+m^2)\omega\big(e(k),\widetilde{e}(k')\big)\,.
    \end{align}
This yields
\begin{align}
\int\dd^dk\,\phi_2(k)\phi_1(k)\delta^d(k^2+m^2)\delta^d(k^2+m^2)(k^2+m^2)\,,
\end{align}
which is indeterminate.
In string theory, appropriate regularization leads to a non-zero contribution.
However, in this toy model, $P$ does not satisfy $P^2=P$, so its interpretation as a projection is subtle.
Although we leave it aside in this work, an important open problem for future study is how to reproduce correct two-point amplitudes without breaking the underlying algebraic structure.
An interesting direction is the approach proposed in \cite{Alfonsi:2024utl}, in which the addition of a boundary term to the action leads to a correct reproduction of the on-shell two-point amplitude in the context of relative $L_\infty$ algebras.
We believe that this approach can be successfully incorporated into our formalism.

For $n\geq 2$, we obtain 
\begin{align}
    &-\lim_{\epsilon\rightarrow 0}\,\omega\Big(Q_\epsilon\Phi_{n+1},\pi_1\bm{f^{(\epsilon)}_0}\bm{h_\epsilon}\bm{D}(Q_\epsilon\Phi_n)\cdots\bm{f^{(\epsilon)}_0}\bm{h_\epsilon}\bm{D}(Q_\epsilon\Phi_1)\bm{1}\Big)\nonumber\\
    =&\lim_{\epsilon\rightarrow 0}\,\omega\big(Q_\epsilon\Phi_{n+1},\pi_1\bm{h_\epsilon m_0f_0^{(\epsilon)}}\bm{h_\epsilon}\bm{D}(Q_\epsilon\Phi_n)\cdots\bm{f^{(\epsilon)}_0}\bm{h_\epsilon}\bm{D}(Q_\epsilon\Phi_1)\bm{1}\big)\nonumber\\
    =&-\lim_{\epsilon\rightarrow 0}\,\omega\big(\Phi_{n+1},\pi_1\bm{m_0f_0^{(\epsilon)}}\bm{h_\epsilon}\bm{D}(Q_\epsilon\Phi_n)\cdots\bm{f^{(\epsilon)}_0}\bm{h_\epsilon}\bm{D}(Q_\epsilon\Phi_1)\bm{1}\big)\,.
\end{align}
If $\Phi_i\in\mathcal{H}^{(P)}_{\mathrm{g.f.}}$, this expression reduces to the minimal model:
\begin{align}
    -\omega\Big(\Phi_{n+1},\pi_1\bm{P}\bm{m_0}\bm{f_0}\bm{P}\pi_n(\Phi_{1}\otimes\cdots\otimes\Phi_n)\Big)\,,\label{eq:minimal model}
\end{align}
which is known as the standard formulation of an on-shell amplitude.
This reduction can be shown by recursive use of the property that $\bm{h_\epsilon D}(Q_\epsilon\Phi_i)$ appends $\Phi_i$ to the rightmost position.
We can rewrite $\bm{f^{(\epsilon)}_0}\bm{h_\epsilon D}(Q_\epsilon\Phi_n)\cdots\bm{f^{(\epsilon)}_0}\bm{h_\epsilon D}(Q_\epsilon\Phi_1)\bm{1}$ recursively as
\begin{align}
    \bm{f^{(\epsilon)}_0}\Bigg(\bm{f^{(\epsilon)}_0}\bigg(\cdots\bm{f^{(\epsilon)}_0}\big(\bm{f^{(\epsilon)}_0}(\Phi_1)\otimes\Phi_2\big)\cdots\otimes\Phi_{n-1}\bigg)\otimes\Phi_n\Bigg)\,,
\end{align}
where each application of $\bm{f_0^{(\epsilon)}}$ is restricted to the terms inside the parentheses.
What we need to show is that for $\Phi_i\in \mathcal{H}^{(P)}_\mathrm{g.f.}$,
\begin{align}
    \bm{f_0}(\Phi_1\otimes\cdots\otimes\Phi_n)=\lim_{\epsilon\rightarrow 0}\bm{f^{(\epsilon)}_0}\Bigg(\bm{f^{(\epsilon)}_0}\bigg(\cdots\bm{f^{(\epsilon)}_0}\big(\bm{f^{(\epsilon)}_0}(\Phi_1)\otimes\Phi_2\big)\cdots\otimes\Phi_{n-1}\bigg)\otimes\Phi_n\Bigg)\,.\label{eq:equivalent minimal model}
\end{align}
From the definition of $\bm{f_0}$, the left-hand side can be written as 
\begin{align}
    \bm{f_0}(\Phi_1\otimes\cdots\otimes\Phi_n)=\sum_{k=0}^\infty \big(-\bm{hm_0}\big)^k\Phi_1\otimes\cdots\otimes\Phi_n\,.
\end{align}
There are two possible ways the action of $\big(-\bm{hm_0}\big)$ can occur: (i) $h$ acts on the rightmost position. (ii) $h$ acts somewhere other than the rightmost position.
In this case, a projection $P$ acts on the rightmost component.
For example, when $\big(\bm{-hM_{0,2}}\big)$ acts on $\Phi_1\otimes\Phi_2\otimes\Phi_3$, the two patterns are:
\begin{align}
   &\mbox{(i)}\qquad -\Phi_1\otimes hM_{0,2}(\Phi_2,\Phi_2)\,,\\
   &\mbox{(ii)}\qquad -hM_{0,2}(\Phi_1,\Phi_2)\otimes P\Phi_3\,.
\end{align}

Since $Ph=0$, once a type (i) operation appears, any subsequent type (ii) operation is forbidden.
As a result, the general structure of $\bm{f_0}(\Phi_1\otimes\cdots\otimes\Phi_n)$ is constrained to the form:
\begin{align}
    \bm{f_0}(\Phi_1\otimes\cdots\otimes\Phi_n)&=\lim_{\epsilon\rightarrow 0}\sum_{k=0}^\infty \sum_{l=0}^\infty \big(-\bm{h_\epsilon m_0}\big)^k\big[\big(-\bm{hm_0}\big)^l(\Phi_1\otimes\cdots\otimes\Phi_{n-1})\otimes P\Phi_n\big]\,,\nonumber\\
    &=\lim_{\epsilon\rightarrow 0}\bm{f^{(\epsilon)}_0}\big(\bm{f_0}(\Phi_1\otimes\cdots\otimes\Phi_{n-1})\otimes P\Phi_n\big)\,,
\end{align}
where the inner $\big(-\bm{hm_0}\big)^l$ acts only within the parentheses, while the presence of $\bm{h_\epsilon}$ ensures that the outer $\big(-\bm{h_\epsilon m_0}\big)^k$ operates only  in type (i).
Applying this recursively, we obtain the desired expression:
\begin{align}
    \bm{f_0}(\Phi_1\otimes\cdots\otimes\Phi_n)=\lim_{\epsilon\rightarrow 0}\bm{f^{(\epsilon)}_0}\Bigg(\bm{f^{(\epsilon)}_0}\bigg(\cdots\bm{f^{(\epsilon)}_0}\big(\bm{f^{(\epsilon)}_0}(\Phi_1)\otimes P\Phi_2\big)\cdots\otimes P\Phi_{n-1}\bigg)\otimes P\Phi_n\Bigg)\,.
\end{align}
This completes the proof that when $\Phi_i\in\mathcal{H}^{(P)}_\mathrm{g.f.}$, the recursive form with $\bm{f_0^{(\epsilon)}}$ reduces to the minimal model.
Consequently, the amplitude formula \eqref{eq:off-shell amplitude with epsilon} derived from the LSZ reduction formula naturally generalizes the standard on-shell amplitude obtained from the minimal model.

\section{BSFT-like action from cohomomorphism}\label{sec:BSFT-like}
In this section, we introduce the cohomomorphism that generates the off-shell amplitude discussed in the previous section and formulate the field redefinition based on it.
The action deformed by this field redefinition closely resembles the action of boundary string field theory (BSFT), which is one of the formulations of open string field theory.

\subsection{Cohomomorphism $\bm{F}$}
Let us consider the cohomomorphism $\bm{F}:\mathcal{TH}\rightarrow\mathcal{TH}$ defined by the following recursive relations:
\begin{subequations}\label{eq:def of F}
\begin{align}
    \bm{F}&=\pi_0+\sum_{n=1}^\infty \big(\pi_1\bm{F}\big)^{\otimes n}\,,\\
    \pi_1\bm{F}&=\pi_1(\bm{I}-\bm{hm_0F})\,.
\end{align}
\end{subequations}
The action of the operator $\pi_1\bm{F}\otimes\pi_1\bm{F}$ follows from the definition \eqref{A otimes B}. Concretely, on $A_1\otimes\cdots\otimes A_n$ it acts as
\begin{align}
    &(\pi_1\bm{F}\otimes\pi_1\bm{F})\big(A_1\otimes\cdots\otimes A_n\big)\nonumber\\
    &=\bigtriangledown(\pi_1\bm{F}\otimes'\pi_1\bm{F})\triangle \big(A_1\otimes\cdots\otimes A_n\big)\,,\nonumber\\
    &=\pi_1\bm{F1}\otimes \pi_1\bm{F}\pi_n(A_1\otimes\cdots\otimes A_n)+\pi_1\bm{F}\pi_n(A_1\otimes\cdots\otimes A_n)\otimes \pi_1\bm{F1}\nonumber\\
    &\quad+\sum_{k=1}^{n-1}\pi_1\bm{F}\pi_k(A_1\otimes\cdots\otimes A_k)\otimes \pi_1\bm{F}\pi_{n-k}(A_{k+1}\otimes\cdots\otimes A_n)\,.\label{eq:tensor product for F}
\end{align}
Note that, because $\pi_1\bm{F}\bm{1}=0$, the first and second terms on the right-hand side vanish.
This action generalizes similarly to higher-order tensor products.
Here, cohomomorphism refers to a map satisfying the co-Leibniz rule with the coproduct:
\begin{align}
    \triangle\bm{F}=(\bm{F}\otimes'\bm{F})\triangle\,.\label{cohomomorphism}
\end{align}
The existence of such a cohomomorphism is not trivial in general; however, at least perturbatively, the above equations \eqref{eq:def of F} can be solved recursively.
Explicitly, we obtain:
\begin{align}
    \pi_1\bm{F}=\pi_1\bm{I}-\pi_1\bm{hm_0}+\pi_1\bm{hm_0}\bm{\mathcal{D}}(\pi_1\bm{hm_0})-\pi_1\bm{hm_0}\bigtriangledown\big(\bm{\mathcal{D}}(\pi_1\bm{hm_0})\otimes'\bm{\mathcal{D}}(\pi_1\bm{hm_0})\big)\triangle+\cdots\,,
\end{align}
where $\bm{\mathcal{D}}(\pi_1\bm{C})$ for a map $\pi_1\bm{C}:\mathcal{TH}\rightarrow\mathcal{H}$ represents the coderivation generated by $\pi_1\bm{C}$, as defined by
\begin{subequations}
    \begin{align}
    &\bm{\mathcal{D}}(\pi_1\bm{C}):=\sum_{n=0}^\infty \bm{\mathcal{D}}(\pi_1\bm{C}\pi_n)\,,\\
    &\bm{\mathcal{D}}(\pi_1\bm{C}\pi_n) (A_1\otimes\cdots\otimes A_m)\nonumber\\
     &:=
   \begin{cases}
        0&\text{if $n>m$}\,,\\
        \pi_1\bm{C}\pi_n(A_1,\dots,A_n)&\text{if $n=m$}\,,\\
        \sum_{k=0}^{m-n}(-1)^{\deg(\bm{C})(\deg(A_1)+\cdots\deg(A_k))}A_1\otimes\cdots\otimes A_k\otimes \pi_1\bm{C}\pi_n(A_{k+1}\otimes\cdots\otimes A_{k+n})\\
        \qquad\qquad\qquad\qquad\qquad\quad\otimes A_{k+n+1}\otimes\cdots\otimes A_m&\text{if $n<m$}\,,
    \end{cases}
\end{align}
\end{subequations}
where $\deg{(\bm{C})}$ denotes the degree of $\pi_1\bm{C}$.
We note that $\bm{\mathcal{D}}(\pi_1\bm{C})=\bm{C}$ when $\bm{C}$ is a coderivation.

This cohomomorphism $\bm{F}$ admits a left inverse $\bm{F}^{-1}$ defined by
\begin{subequations}\label{eq:def of F inv}
\begin{align}
    \bm{F}^{-1}&:=\pi_0+\sum_{n=1}^\infty \big(\pi_1\bm{F}^{-1}\big)^{\otimes n}\,,\\
    \pi_1\bm{F}^{-1}&:=\pi_1(\bm{I}+\bm{hm_0})\,.
\end{align}
\end{subequations}
It is also a cohomomorphism.
The tensor product of the operators $\pi_1\bm{F}^{-1}$ is defined in the same way as in \eqref{eq:tensor product for F}.
To verify that $\bm{F}^{-1}$ is indeed a left inverse of $\bm{F}$, observe that
\begin{align}
    \pi_1\bm{F}^{-1}\bm{F}&=\pi_1(\bm{I}+\bm{hm_0})\bm{F}\,,\nonumber\\
    &=\pi_1\bm{F}+\pi_1\bm{hm_0F}\,,\nonumber\\
    &=\pi_1(\bm{I}-\bm{hm_0F}+\bm{hm_0F})\,,\nonumber\\
    &=\pi_1\bm{I}\,.
\end{align}
Here, the first equality follows from the definition of $\bm{F}^{-1}$ \eqref{eq:def of F inv}, and the third equality uses the definition of $\bm{F}$ \eqref{eq:def of F}.
Since the cohomomorphism is fully determined by its $\pi_1$ component, this is sufficient to establish left invertibility.
Conversely, demonstrating that $\bm{F}^{-1}$ is also a right inverse is less trivial.
In general, utilizing the definitions \eqref{eq:def of F} and \eqref{eq:def of F inv}, we compute $\pi_1\bm{F}\bm{F}^{-1}$ as follows:
\begin{align}
    \pi_1\bm{F}\bm{F}^{-1}&=\pi_1(\bm{I}-\bm{hm_0F})\bm{F}^{-1}\,,\nonumber\\
    &=\pi_1\bm{F}^{-1}-\pi_1\bm{hm_0FF}^{-1}\,,\nonumber\\
    &=\pi_1\bm{I}+\pi_1\bm{hm_0}-\pi_1\bm{hm_0}\bm{FF}^{-1}\,.
\end{align}
The resulting recursive expression makes it non-trivial whether the second and third terms on the right-hand side cancel out. However, within the framework of perturbation theory, solving this expression recursively reveals that the second and third terms cancel at every order.
Therefore, within perturbation theory, $\bm{F}^{-1}$ acts as both the left and right inverse of $\bm{F}$.
In the following, we assume this perturbative behavior and proceed under the assumption that $\bm{F}$ exists, and $\bm{F}^{-1}$ satisfies both: 
\begin{align}
    \bm{F}^{-1}\bm{F}=\bm{FF}^{-1}=\bm{\bm{I}}\,.
\end{align}

An important property of $\bm{F}$ is that it generates the off-shell amplitude.
Specifically, we have:
\begin{align}
    \pi_1\bm{m_0F}(\Phi_1\otimes\cdots\otimes\Phi_n)=\lim_{\epsilon\rightarrow 0}\pi_1\bm{m_0f^{(\epsilon)}_0}\bm{h_\epsilon}\bm{D}(Q_\epsilon\Phi_n)\cdots\bm{f^{(\epsilon)}_0}\bm{h_\epsilon}\bm{D}(Q_\epsilon\Phi_1)\bm{1}.\label{eq:mF=off-shell amp}
\end{align}
This relation ensures that $\bm{F}$ correctly reproduces the off-shell amplitude discussed in the previous section.
This equality can be proven by mathematical induction, showing that:
\begin{align}
    \lim_{\epsilon\rightarrow 0}\bm{F}^{-1}\bm{f^{(\epsilon)}_0}\bm{h_\epsilon D}(Q_\epsilon\Phi_n)\cdots\bm{f^{(\epsilon)}_0}\bm{h_\epsilon D}(Q_\epsilon\Phi_1)\bm{1}=\Phi_1\otimes\cdots\otimes\Phi_n\,,\label{eq:F=fhD}
\end{align}
for degree-even elements $\Phi_i\in \mathcal{H}_\mathrm{g.f.}$.
For this proof, we use the following identities:
\begin{align}
    \triangle \bm{f^{(\epsilon)}_0}&=\big[\bm{f^{(\epsilon)}_0}\otimes'\pi_0+\bm{I}\otimes'\bm{f^{(\epsilon)}_0}\big]\triangle\,,\label{eq:formula for f}\\
    \triangle \bm{h_\epsilon D}(Q_\epsilon \Phi)&=\big[\bm{h_\epsilon D}(Q_\epsilon\Phi)\otimes'\pi_0+\bm{I}\otimes'\bm{h_\epsilon D}(Q_\epsilon\Phi)\big]\triangle\,,\label{eq:formula for D'}
\end{align}
The first identity holds when acting on the tensor algebra of the subspace annihilated by $h_\epsilon$.
These identities are straightforward consequences of the definitions.
Here, we present another formula to be used later:
\begin{align}
\lim_{\epsilon\to 0}\,\pi_1\bm{F}^{-1}\bm{f}_0^{(\epsilon)}=\lim_{\epsilon\to 0}\pi_1\,.\label{formula pi1}
\end{align}
This can be verified as follows.
Writing $\bm{F}^{-1}$ and $\bm{f}_0^{(\epsilon)}$ according to their definitions, the left-hand side becomes
\begin{align}
\lim_{\epsilon\to 0}\,\pi_1(\bm{I}+\bm{h}\bm{m}_0)\frac{1}{\bm{I}+\bm{h}_\epsilon\bm{m}_0}.
\end{align}
As mentioned earlier, $\bm{h}$ and $\bm{h}_\epsilon$ remain different operators even in the limit $\epsilon\to 0$.
However, $\pi_1\bm{h}=h$ and $\pi_1\bm{h}_\epsilon$=$h_\epsilon$ coincide.
Therefore, in this case,
\begin{align}
\lim_{\epsilon\to 0}\,\pi_1(\bm{I}+\bm{h}\bm{m}_0)\frac{1}{\bm{I}+\bm{h}_\epsilon\bm{m}_0}
=
\lim_{\epsilon\to 0}\,\pi_1(\bm{I}+\bm{h}_\epsilon\bm{m}_0)\frac{1}{\bm{I}+\bm{h}_\epsilon\bm{m}_0}
=
\lim_{\epsilon\to 0}\pi_1 .
\end{align}
Let us begin with the case $n=1$.
By using the identity \eqref{eq:pi_n identity}, the $\pi_k$ component of the expression is computed as
\begin{align}
    \pi_k \bm{F}^{-1}\bm{f^{(\epsilon)}_0}\bm{h_\epsilon D}(Q_\epsilon\Phi_1)\bm{1}&=\bigtriangledown(\pi_{k-1}\otimes'\pi_1)\triangle \bm{F}^{-1}\bm{f^{(\epsilon)}_0}\bm{h_\epsilon D}(Q_\epsilon\Phi_1)\bm{1}\,.
\end{align}
Applying the identities stated above, this expands to
\begin{align}
   \lim_{\epsilon\to 0}\Big[ &\pi_{k-1}\bm{F}^{-1}\bm{f^{(\epsilon)}_0}\bm{h_\epsilon D}(Q_\epsilon\Phi_1)\bm{1}\otimes\pi_1\bm{F}^{-1}\bm{1}+\pi_{k-1}\bm{F}^{-1}\bm{h_\epsilon}\bm{D}(Q_\epsilon \Phi_1)\bm{1}\otimes \pi_1\bm{F}^{-1}\bm{f_0^{(\epsilon)}}\bm{1}\nonumber\\
    &+\pi_{k-1}\bm{F}^{-1}\bm{1}\otimes\pi_1\bm{F}^{-1}\bm{f_0^{(\epsilon)}}\bm{h_\epsilon D}(Q_\epsilon\Phi_1)\bm{1}\Big]\,.
\end{align}
Here, we also used the fact that $\bm{F}^{-1}$ is a cohomomorphism, which means that the identity $\triangle\bm{F}^{-1}=\big(\bm{F}^{-1}\otimes'\bm{F}^{-1}\big)\triangle$ holds.
The first term vanishes because $\bm{F}^{-1}\bm{1}=\bm{1}$.
The second term also vanishes by \eqref{formula pi1}.
The third term survives only for $k=1$, and by \eqref{formula pi1} one finds
\begin{align}
\lim_{\epsilon\to 0}\pi_1\bm{F}^{-1}\bm{f^{(\epsilon)}_0}\bm{h_\epsilon D}(Q_\epsilon\Phi_1)\bm{1}=\lim_{\epsilon\to 0}\pi_1\bm{h}_\epsilon\bm{D}(Q_\epsilon \phi_1)\bm{1}=\Phi_1.
\end{align}
Thus, we obtain
\begin{align}
    \lim_{\epsilon\to 0}\bm{F}^{-1}\bm{f^{(\epsilon)}_0}\bm{h_\epsilon D}(Q_\epsilon\Phi_1)\bm{1}=\Phi_1\,.
\end{align}
Now, assume that the statement \eqref{eq:F=fhD} holds for all $n\leq m$, and consider the case $n=m+1$.
Applying the identities \eqref{eq:formula for f} and \eqref{eq:formula for D'}, we obtain:
\begin{align}
    &\lim_{\epsilon\rightarrow 0}\triangle \bm{F}^{-1}\bm{f^{(\epsilon)}_0}\bm{h_\epsilon D}(Q_\epsilon\Phi_{m+1})\cdots\bm{f^{(\epsilon)}_0}\bm{h_\epsilon D}(Q_\epsilon\Phi_1)\bm{1}\nonumber\\
    &=\lim_{\epsilon\rightarrow 0}\Big[\bm{1}\otimes'\bm{F}^{-1}\bm{f^{(\epsilon)}_0}\bm{h_\epsilon D}(Q_\epsilon\Phi_{m+1})\cdots\bm{f^{(\epsilon)}_0}\bm{h_\epsilon D}(Q_\epsilon\Phi_1)\bm{1}\nonumber\\
    &\quad+\bm{F}^{-1}\bm{f^{(\epsilon)}_0}\bm{h_\epsilon D}(Q_\epsilon\Phi_{m+1})\cdots\bm{f^{(\epsilon)}_0}\bm{h_\epsilon D}(Q_\epsilon\Phi_1)\bm{1}\otimes'\bm{1}\nonumber\\
    &\quad+\sum_{l=1}^{m}\bm{F}^{-1}\bm{f^{(\epsilon)}_0}\bm{h_\epsilon D}(Q_\epsilon\Phi_{m-l+1})\cdots\bm{f^{(\epsilon)}_0}\bm{h_\epsilon D}(Q_\epsilon\Phi_1)\bm{1}\nonumber\\
    &\quad\qquad\otimes'\bm{F}^{-1}\bm{f^{(\epsilon)}_0}\bm{h_\epsilon D}(Q_\epsilon\Phi_{m+1})\cdots\bm{f^{(\epsilon)}_0}\bm{h_\epsilon D}(Q_\epsilon\Phi_{m-l+2})\bm{1}\Big]\,.\label{eq:calculation}
\end{align}
By the induction hypothesis, the third term reduces to:
\begin{align}
    \sum_{l=1}^m\Phi_1\otimes\cdots\otimes\Phi_{m-l+1}\otimes'\Phi_{m-l+2}\otimes\cdots\otimes\Phi_{m+1}\,.
\end{align}
We consider applying $\bigtriangledown(\pi_{k-1}\otimes'\pi_1)$ from the left here.
First, the right tensor factor in the first term of the expression $\eqref{eq:calculation}$ vanishes, where the expression is given by
\begin{align}
    \lim_{\epsilon\to 0}\pi_1\bm{F}^{-1}\bm{f^{(\epsilon)}_0}\bm{h_\epsilon D}(Q_\epsilon\Phi_{m+1})\cdots\bm{f^{(\epsilon)}_0}\bm{h_\epsilon D}(Q_\epsilon\Phi_1)\bm{1}\,.
\end{align}
This vanishes due to the following reasoning.
The operator $\bm{h_\epsilon D}(Q_\epsilon\Phi_{m+1})$ is defined such that it increases the number of tensor factors by one.
Since the operator preceding it already contains at least one tensor factor, the entire expression
\begin{align}
    \bm{h_\epsilon D}(Q_\epsilon\Phi_{m+1})\cdots\bm{f^{(\epsilon)}_0}\bm{h_\epsilon D}(Q_\epsilon\Phi_1)\bm{1}\,,
\end{align}
necessarily has two or more tensor factors.
Therefore, by \eqref{formula pi1}, this expression vanishes.
Furthermore, the second term of \eqref{eq:calculation} trivially vanishes because $\pi_1\bm{1}=0$, leaving only the third term.
Regarding the third term, all contributions except the $l=1$ component also vanish. Consequently, we obtain the following expression:
\begin{align}
    \lim_{\epsilon\rightarrow 0}\pi_{k}\bm{F}^{-1}\bm{f^{(\epsilon)}_0}\bm{h_\epsilon D}(Q_\epsilon\Phi_{m+1})\cdots\bm{f^{(\epsilon)}_0}\bm{h_\epsilon D}(Q_\epsilon\Phi_1)\bm{1}
    =\bigtriangledown\big(\pi_{k-1}(\Phi_1\otimes\cdots\otimes\Phi_{m})\otimes'\pi_1\Phi_{m+1}\big)\,.
\end{align}
Since the expression yields $\Phi_1\otimes\cdots\otimes\Phi_{m+1}$ exclusively when $k=m+1$, and vanishes for all other values of $k$, we conclude:
\begin{align}
    \lim_{\epsilon\rightarrow 0}\pi_{k}\bm{F}^{-1}\bm{f^{(\epsilon)}_0}\bm{h_\epsilon D}(Q_\epsilon\Phi_{m+1})\cdots\bm{f^{(\epsilon)}_0}\bm{h_\epsilon D}(Q_\epsilon\Phi_1)\bm{1} =
    \begin{cases}
        \Phi_1\otimes\cdots\otimes\Phi_{m+1}&\text{if $k=m+1$}\,,\\
        0&\text{otherwise}\,.
    \end{cases}
\end{align}
This completes the inductive step and proves that:
\begin{align}
     \lim_{\epsilon\rightarrow 0}\bm{F}^{-1}\bm{f^{(\epsilon)}_0}\bm{h_\epsilon D}(Q_\epsilon\Phi_{m+1})\cdots\bm{f^{(\epsilon)}_0}\bm{h_\epsilon D}(Q_\epsilon\Phi_1)\bm{1}=\Phi_1\otimes\cdots\otimes\Phi_{m+1}\,.
\end{align}
Applying $\pi_1\bm{m_0F}$ to this result, we recover the off-shell amplitude \eqref{eq:mF=off-shell amp}.
Since the LSZ-reduced expression \eqref{eq:off-shell amplitude with epsilon} coincides with the minimal model when all inputs belong to $\mathcal{H}^{(P)}_\mathrm{g.f.}$, it follows that $\pi_1\bm{m_0F}$ also reproduces the minimal model in that case.
In particular, on $\mathcal{H}_\mathrm{g.f.}$, we have
\begin{align}
    \bm{FP}=\bm{f_0P}\,.\label{eq:on-shell F}
\end{align}

Here, we introduce two important properties of $\bm{F}$.
The first property is that $\pi_1\bm{m_0F}$ is cyclic:
\begin{align}
   \omega\big(\Phi_1,\pi_1\bm{m_0F}(\Phi_2\otimes\cdots\otimes\Phi_{n+1})\big)=\omega\big(\Phi_{n+1},\pi_1\bm{m_0F}(\Phi_1\otimes\cdots\otimes\Phi_{n})\big)\,,\label{eq:mF cyclicity}
\end{align}
This implies that the off-shell amplitude is cyclic.
To prove this, it suffices to show that 
\begin{align}
    \bra{\omega}\big(\pi_1\bm{m_0F}\otimes\pi_1+\pi_1\otimes\pi_1\bm{m_0F}\big)=0\,.\label{eq:pi_2mF}
\end{align}
The operators $\pi_1\bm{m_0F}\otimes\pi_1$ and $\pi_1\otimes\pi_1\bm{m_0F}$ are to be understood, according to definition \eqref{A otimes B}, as maps on $\mathcal{TH}$.
Concretely, they act on the tensor product $A_1\otimes\cdots\otimes A_n$ as follows:
\begin{align}
    &\bra{\omega}\big(\pi_1\bm{m_0F}\otimes\pi_1+\pi_1\otimes\pi_1\bm{m_0F}\big)\big(A_1\otimes\cdots\otimes A_n\big)\nonumber\\
    &=\bra{\omega}\bigtriangledown\big(\pi_1\bm{m_0F}\otimes'\pi_1+\pi_1\otimes'\pi_1\bm{m_0F}\big)\triangle\big(A_1\otimes\cdots\otimes A_n\big)\,,\nonumber\\
    &=\bra{\omega}\big(\pi_1\bm{m_0F}\pi_{n-1}(A_1\otimes\cdots\otimes A_{n-1})\otimes A_n+(-1)^{\deg(A_1)}A_1\otimes\pi_1\bm{m_0F}\pi_{n-1}(A_2\otimes\cdots\otimes A_n)\big)\,,\nonumber\\
    &=\omega\big(\pi_1\bm{m_0F}\pi_{n-1}(A_1\otimes\cdots\otimes A_{n-1}), A_n\big)+(-1)^{\deg(A_1)}\omega\big(A_1,\pi_1\bm{m_0F}\pi_{n-1}(A_2\otimes\cdots\otimes A_n)\big)\,.
\end{align}
When the anti-symmetric property of $\omega$ \eqref{eq:anti-sym omega} is applied, the expression directly becomes the condition for cyclicity.
Utilizing the cohomomorphism property of $\bm{F}$, given by \eqref{cohomomorphism}, this expression can be written as
\begin{align}
    \bra{\omega}\big(\pi_1\bm{m_0F}\otimes\pi_1+\pi_1\otimes\pi_1\bm{m_0F}\big)=\bra{\omega}\big(\pi_1\bm{m_0}\otimes\pi_1\bm{F}^{-1}+\pi_1\bm{F}^{-1}\otimes\pi_1\bm{m_0}\big)\bm{F}\,.
\end{align}
Here, as maps on $\mathcal{TH}$, $\pi_1\bm{m}_0\otimes\pi_1\bm{F}^{-1}$ and $\pi_1\bm{F}^{-1}\otimes\pi_1\bm{m}_0$ are understood according to the definition in \eqref{A otimes B}.
Furthermore, utilizing the definition of $\bm{F}^{-1}$ \eqref{eq:def of F inv}, this expression can be written as
\begin{align}
    \bra{\omega}\big(\pi_1\bm{m_0F}\otimes\pi_1+\pi_1\otimes\pi_1\bm{m_0F}\big)
&=\bra{\omega}\big(\pi_1\bm{m_0}\otimes\pi_1+\pi_1\otimes\pi_1\bm{m_0}\big)\bm{F}\nonumber\\
    &\quad+\bra{\omega}\big(\pi_1\bm{m_0}\otimes h\pi_1\bm{m_0}+h\pi_1\bm{m_0}\otimes\pi_1\bm{m_0}\big)\bm{F}\,.
\end{align}
The first term on the right-hand side vanishes due to cyclicity of $\bm{m_0}$, and the second term vanishes because of the assumed property of $h$ \eqref{eq:h property}.

The second property is the following relation:
\begin{align}
    \pi_1\bm{F}^{-1}\bm{M_0F}=\pi_1\bm{Q}+\pi_1\bm{Pm_0F}\,.\label{eq:pi_1F^-1MF}
\end{align}
This can be verified as follows.
First, by using the definition of $\bm{F}^{-1}$ \eqref{eq:def of F inv} and splitting $\bm{M}_0$ into $\bm{Q}$ and $\bm{m}_0$, the left-hand side can be written as 
\begin{align}
    \pi_1\bm{F}^{-1}\bm{M_0F}&=\pi_1(\bm{I}+\bm{hm_0})(\bm{Q}+\bm{m_0})\bm{F}\,,\nonumber\\
    &=\pi_1\bm{QF}+\pi_1\bm{m_0F}+\pi_1\bm{hm_0}(\bm{Q}+\bm{m_0})\bm{F}\,.\label{calculation2}
\end{align}
The first term on the right-hand side can be written as
\begin{align}
    \pi_1\bm{QF}&=Q\pi_1\bm{F}\,,\nonumber\\
    &=Q\pi_1(\bm{I}-\bm{hm_0F})\,,\nonumber\\
    &=\pi_1\bm{Q}-\pi_1\bm{Qhm_0F}\,,
\end{align}
by utilizing the identity $\pi_1\bm{Q}=Q\pi_1$ and the definition of $\bm{F}$ \eqref{eq:def of F}.
Furthermore, the third term in \eqref{calculation2} can be written as
\begin{align}
    \pi_1\bm{hm_0}(\bm{Q}+\bm{m_0})\bm{F}=-\pi_1\bm{hQm_0F}\,,
\end{align}
by applying the $A_\infty$ relations \eqref{eq:A infty relations}.
Combining these results and applying the identity $\bm{Qh}+\bm{hQ}=\bm{I}-\bm{P}$, we ultimately obtain \eqref{eq:pi_1F^-1MF}.
The operator $\bm{F}^{-1}\bm{M}_0\bm{F}$ is a coderivation. This can be seen immediately by applying the coproduct $\triangle$ from the left:
\begin{align}
    \triangle\bm{F}^{-1}\bm{M_0F}&=(\bm{F}^{-1}\otimes'\bm{F}^{-1})(\bm{I}\otimes'\bm{M_0}+\bm{M_0}\otimes'\bm{I})(\bm{F}\otimes'\bm{F})\triangle\,,\nonumber\\
    &=\Big(\bm{I}\otimes'\bm{F}^{-1}\bm{M_0F}+\bm{F}^{-1}\bm{M_0F}\otimes'\bm{I}\Big)\triangle\,.
\end{align}
This is the definition of a coderivation.
A coderivation is fully determined by its $\pi_1$ component.
Therefore, \eqref{eq:pi_1F^-1MF} holds as a coderivation:
\begin{align}
   \bm{F}^{-1}\bm{M_0F}=\bm{Q}+\bm{\mathcal{D}}(\pi_1\bm{Pm_0F})\,.\label{eq:F^-1MF coder}
\end{align}
Since $\bm{M_0}$ is nilpotent, the left-hand side of \eqref{eq:F^-1MF coder} also squares to zero.
Therefore, the cohomomorphism $\bm{F}$ (along with its inverse $\bm{F}^{-1}$) induces a new $A_\infty$ algebra. 

\subsection{Field redefinition by using $\bm{F}$}

Let us consider the following field redefinition using the cohomomorphism $\bm{F}$,
\begin{align}
    \Phi=\pi_1\bm{F}\frac{1}{1-\Psi}\,,\label{eq:Phi redef}
\end{align}
where $\Psi$ is the redefined field, which can be expanded as 
\begin{align}
    \Psi:= \sum_{e_a\in\mathcal{H}_\mathrm{g.f.}}\psi^ae_a\,,
\end{align}
and $\frac{1}{1-\Psi}$ is defined by 
\begin{align}
    \frac{1}{1-\Psi}:=\bm{1}+\sum_{n=1}^\infty \Psi^{\otimes n}\,.\label{eq:def of group-like element}
\end{align}
The quantity $\frac{1}{1-\Psi}$ satisfies the identity with respect to the product $\triangle$:
\begin{align}
    \triangle\frac{1}{1-\Psi}=\frac{1}{1-\Psi}\otimes'\frac{1}{1-\Psi}\,.\label{group-like element}
\end{align}
Such a quantity is called a group-like element.
Since $\bm{F}$ is a cohomomorphism, for group-like element $\frac{1}{1-\Phi}=\bm{1}+\sum_{n=1}^\infty \Phi^{\otimes n}$, the equation
\begin{align}
    \frac{1}{1-\Phi}=\bm{F}\frac{1}{1-\Psi}\,,\label{eq:Phi = F psi}
\end{align}
holds.
For instance, applying $\pi_2=\bigtriangledown(\pi_1\otimes'\pi_1)\triangle$ from the left to the right-hand side, and utilizing the cohomomorphism property \eqref{cohomomorphism} and the group-like element property \eqref{group-like element}, the expression can be written as
\begin{align}
    \pi_1\bm{F}\frac{1}{1-\Psi}\otimes \pi_1\bm{F}\frac{1}{1-\Psi}\,.
\end{align}
This exactly matches the result $\pi_2\frac{1}{1-\Phi}=\Phi\otimes\Phi$.
The same can be verified for higher-order components.
Conversely, it implies that $\Psi$ can be written in terms of $\Phi$ as
\begin{align}
    \Psi=\pi_1\bm{F}^{-1}\frac{1}{1-\Phi}\,.
\end{align}
This expression can also be rewritten as follows.
First, utilizing the definition of $\bm{F}^{-1}$ \eqref{eq:def of F inv} and $\bm{M}_0=\bm{Q}+\bm{m}_0$, we can write $\Phi$ as
\begin{align}
    \Phi=\pi_1\big(\bm{I}-\bm{hQ}+\bm{hM_0}\big)\frac{1}{1-\Psi}\,.
\end{align}
Furthermore, by employing the identity $\bm{Qh}+\bm{hQ}=\bm{I}-\bm{P}$, we can replace $\bm{I}-\bm{hQ}$ with $\bm{Qh}+\bm{P}$, and the expression becomes
\begin{align}
    \Phi=\pi_1\big(\bm{Qh}+\bm{P}+\bm{hM_0}\big)\frac{1}{1-\Psi}=Qh\Phi+P\Phi+\pi_1\bm{hM_0}\frac{1}{1-\Psi}\,.
\end{align}
The first term, $Qh\Phi$, vanishes because of $h\Phi=0$.
This ultimately yields
\begin{align}
    \Psi=P\Phi+\pi_1\bm{hM_0}\frac{1}{1-\Phi}\,.\label{eq:field redef Psi}
\end{align}
From this expression, the relationship between $\Phi$ and the redefined field $\Psi$ becomes clear.
For a solution of the equation of motion $\pi_1\bm{M_0}\frac{1}{1-\Phi}=0$ for $\Phi$, the corresponding $\Psi$ belongs to the $Q$-cohomology.
When the $Q$-cohomology is trivial, such as in bosonic open string field theory on the tachyon vacuum, the corresponding $\Psi$ for any solution of the equation of motion for $\Phi$ is always zero.
This might seem strange because, in this case, $\Psi$ cannot distinguish between trivial and nontrivial solutions.
However, this issue stems from the subtlety of the existence of $\bm{F}$ and the fact that $\bm{F}^{-1}$ being a right inverse of $\bm{F}$ is non-trivial in a non-perturbative setting.
Since we are considering only the perturbation theory in this work, this issue does not pose a problem.
In general, even if $\Psi$ belongs to the $Q$-cohomology, the corresponding $\Phi$, defined by equation \eqref{eq:Phi redef}, does not necessarily solve the equation of motion $\pi_1\bm{M_0}\frac{1}{1-\Phi}=0$.
The equation of motion for $\Phi$ is equivalent to the following equation expressed in terms of $\Psi$:
\begin{align}
    \pi_1\bm{F}^{-1}\bm{M_0F}\frac{1}{1-\Psi}=0\,.
\end{align}
By using \eqref{eq:F^-1MF coder}, this equation can be written as
\begin{align}
    Q\Psi+\pi_1\bm{Pm_0F}\frac{1}{1-\Psi}=0\,.
\end{align}
Since the projection operators $P$ and $\mathbb{I}-P$ act independently, this equality requires that each term vanishes separately:
\begin{align}
    Q\Psi=0,\qquad \pi_1\bm{Pm_0F}\frac{1}{1-\Psi}=0\,.
\end{align}
Given that $\Psi\in \mathcal{H}_\mathrm{g.f.}$, the first equation implies that $\Psi$ belongs to the $Q$-cohomology.
The second equation then imposes an additional condition: not only must $\Psi$ belong to $Q$-cohomology, but it must also yield a vanishing on-shell amplitude.

\subsection{BSFT-like action}
Let us rewrite the tree-level Maurer-Cartan action:
\begin{align}
    S_0=-\sum_{n=1}^{\infty} \frac{1}{n+1}\omega\Big(\Phi,\pi_1\bm{M_{0,n}}\frac{1}{1-\Phi}\Big)\,,\label{eq:original action}
\end{align}
using the field redefinition as:
\begin{align}
    S_0=-\sum_{n=1}^\infty \frac{1}{n+1}\omega\Big(\pi_1\bm{F}\frac{1}{1-\Psi},\pi_1\bm{M_{0,n}}\bm{F}\frac{1}{1-\Psi}\Big)\,.\label{eq:action in terms of Psi}
\end{align}
For computational convenience, we employ the following equivalent expression:
\begin{align}
    S_0=-\int_0^1\dd t\,\omega\Big(\pdv{}{t}\pi_1\bm{F}\frac{1}{1-\Psi(t)},\pi_1\bm{M_0F}\frac{1}{1-\Psi(t)}\Big)\,,\label{eq:action expression}
\end{align}
where $\Psi(t)$ is a smooth function satisfying:
\begin{align}
    \Psi(1)=\Psi,\qquad \Psi(0)=0\,.
\end{align}
Since the left input of $\omega$ in \eqref{eq:action expression} is annihilated by $h$, we can insert $\bm{F}^{-1}$ into the right input and apply \eqref{eq:pi_1F^-1MF} to rewrite the expression as follows:
\begin{align}
   S_0=-\int_0^1\dd t\,\omega\Big(\pdv{}{t}\pi_1\bm{F}\frac{1}{1-\Psi(t)},Q\Psi(t)\Big)-\int_0^1\dd t\,\omega\Big(\pdv{}{t}\pi_1\bm{F}\frac{1}{1-\Psi(t)},\pi_1\bm{Pm_0F}\frac{1}{1-\Psi(t)}\Big)\,.
\end{align}
Let $\bm{D}\big(\pdv{}{t}\Psi(t)\big)$ denote the coderivation generated by $\pdv{}{t}\Psi(t)$ as defined in \eqref{eq:def of D}.
Then, \break $\pdv{}{t}\pi_1\bm{F}\frac{1}{1-\Psi(t)}$ is written as
\begin{align}
    \pdv{}{t}\pi_1\bm{F}\frac{1}{1-\Psi(t)}=\pi_1\bm{F}\bm{D}\big(\pdv{}{t}\Psi(t)\big)\frac{1}{1-\Psi(t)}\,.
\end{align}
Substituting these into the action, we have:
\begin{align}
    S_0=&-\int_0^1\dd t\,\omega\Big(\pi_1\bm{F}\bm{D}\big(\pdv{}{t}\Psi(t)\big)\frac{1}{1-\Psi(t)},Q\Psi(t)\Big)\,,\\
    &-\int_0^1\dd t\,\omega\Big(\pi_1\bm{F}\bm{D}\big(\pdv{}{t}\Psi(t)\big)\frac{1}{1-\Psi(t)},\pi_1\bm{Pm_0F}\frac{1}{1-\Psi(t)}\Big)\,.
\end{align}
For the second term, using $P\pi_1\bm{F}=P$, we rewrite it as:
\begin{align}
    -\int_0^1\dd t\,\omega\Big(\pdv{}{t}\Psi(t),\pi_1\bm{Pm_0F}\frac{1}{1-\Psi(t)}\Big)\,.\label{eq:tree action calculation}
\end{align}
We now compute the first term using the definition of $\bm{F}$ \eqref{eq:def of F}:
\begin{align}
    &-\int_0^1\dd t\,\omega\Big(\pi_1\bm{F}\bm{D}\big(\pdv{}{t}\Psi(t)\big)\frac{1}{1-\Psi(t)},Q\Psi(t)\Big)\nonumber\\
    &=-\int_0^1\dd t\,\omega\Big(\pdv{}{t}\Psi(t),Q\Psi(t)\Big)+\int_0^1\dd t\,\omega\Big(\pi_1\bm{hm_0F}\bm{D}\big(\pdv{}{t}\Psi(t)\big)\frac{1}{1-\Psi(t)},Q\Psi(t)\Big)\nonumber\\
    &=-\int_0^1\dd t\,\omega\Big(\pdv{}{t}\Psi(t),Q\Psi(t)\Big)+\int_0^1\dd t\,\omega\Big(hQ\Psi(t),\pi_1\bm{m_0F}\bm{D}\big(\pdv{}{t}\Psi(t)\big)\frac{1}{1-\Psi(t)}\Big)\,,
\end{align}
where we used the property of $h$ \eqref{eq:h property} and the anti-symmetric property of $\omega$ \eqref{eq:anti-sym omega} at the second equality.
Applying $Qh+hQ=\mathbb{I}-P$, the second term becomes:
\begin{align}
    \int_0^1\dd t\,\omega\Big((\mathbb{I}-P)\Psi(t),\pi_1\bm{m_0F}\bm{D}\big(\pdv{}{t}\Psi(t)\big)\frac{1}{1-\Psi(t)}\Big)\,.
\end{align}
Combining all terms and inserting the trivial identity
\begin{align}
    0=-\int_0^1\dd t\,\omega\Big(\pdv{}{t}\Psi(t),\pi_1\bm{m_0F}\frac{1}{1-\Psi(t)}\Big)+\int_0^1\dd t\,\omega\Big(\pdv{}{t}\Psi(t),\pi_1\bm{m_0F}\frac{1}{1-\Psi(t)}\Big),
\end{align}
we obtain:
\begin{align}
    S_0=&-\int_0^1\dd t\,\omega\Big(\pdv{}{t}\Psi(t),Q\Psi(t)\Big)-\int_0^1\dd t\,\omega\Big(\pdv{}{t}\Psi(t),\pi_1\bm{m_0F}\frac{1}{1-\Psi(t)}\Big)\nonumber\\
    &+\int_0^1\dd t\,\omega\Big((\mathbb{I}-P)\pdv{}{t}\Psi(t),\pi_1\bm{m_0F}\frac{1}{1-\Psi(t)}\Big)\nonumber\\
    &+\int_0^1\dd t\,\omega\Big((\mathbb{I}-P)\Psi(t),\pi_1\bm{m_0F}\bm{D}\big(\pdv{}{t}\Psi(t)\big)\frac{1}{1-\Psi(t)}\Big)\,.
\end{align}
Thanks to the cyclicity of $\pi_1\bm{m_0F}$, which we have seen in the previous subsection, the integrand can be expressed as a total derivative with respect to $t$.
This allows us to explicitly perform the $t$-integration and obtain the following expression:
\begin{align}
    S_0=-\frac{1}{2}\omega\big(\Psi,Q\Psi\big)-\sum_{n=2}^\infty \frac{1}{n+1}\omega\Big(\Psi,\pi_1\bm{m_0F}\pi_n\frac{1}{1-\Psi}\Big)+\omega\Big((\mathbb{I}-P)\Psi,\pi_1\bm{m_0F}\frac{1}{1-\Psi}\Big)\,.
\end{align}
For clarity, let us define:
\begin{align}
    \pi_1\bm{W}:=-Q+\pi_1\bm{m_0F}\,,
\end{align}
which is cyclic.
As we have seen, the first term corresponds to the two-point amplitude, while the second term encodes the higher-point amplitudes.
Note that the relative sign is opposite.
The tree-level action can now be compactly written as:
\begin{align}
    S_0=-\sum_{n=1}^\infty \frac{1}{n+1}\omega\Big(\Psi,\pi_1\bm{W}\pi_n\frac{1}{1-\Psi}\Big)+\omega\Big((\mathbb{I}-P)\Psi,\pi_1\bm{W}\frac{1}{1-\Psi}\Big)\,.\label{eq:BSFT-like action}
\end{align}
The identity $(\mathbb{I}-P)Q=Q$ was used in the simplification of this action.
We refer to this as the \textit{BSFT-like action}.
The naming reflects the structural similarity to the Boundary String Field Theory (BSFT) action.
The BSFT is one of the formulations of open string field theory \cite{Witten:1992qy,Witten:1992cr,Shatashvili:1993kk,Shatashvili:1993ps,Li:1993za}, and is given by:
\begin{align}
    S_{\mathrm{BSFT}}=Z_{D_2}-\sum_a\beta^a\pdv{}{\psi^a}Z_{D_2}\,,\label{eq:BSFT action}
\end{align}
where, $Z_{D_2}$ is the disk partition function of worldsheet QFT, and $\beta^a$ are the worldsheet $\beta$-functions.
In the following subsections, we compare the BSFT-like action \eqref{eq:BSFT-like action} with the BSFT action \eqref{eq:BSFT action}, highlighting their similarities and differences.

\subsection{$A_\infty$ structure of BSFT-like action}
We begin by examining the $A_\infty$ structure of the BSFT-like action \eqref{eq:BSFT-like action}, and comparing it with that of the BSFT action.
The expression of the BSFT-like action \eqref{eq:BSFT-like action} does not take the form of a Maurer–Cartan action.
Therefore, the underlying algebraic structure is not immediately evident.
In the original formulation of BSFT \cite{Witten:1992qy}, it is not the BSFT action $S_{\mathrm{BSFT}}$ itself but rather the differential $\dd S_{\mathrm{BSFT}}$ that is constructed algebraically based on the BV formalism.
Motivated by this, we examine the differential of the BSFT-like action: \begin{align}
    \dd S_0=-\dd\sum_{n=1}^\infty \frac{1}{n+1}\omega\Big(\pi_1\bm{F}\frac{1}{1-\Psi},\pi_1\bm{M_{0,n}F}\frac{1}{1-\Psi}\Big)\,.
\end{align}
This can be rewritten as:
\begin{align}
    \dd S_0=-\omega\Big(\pi_1\bm{F}\bm{D}(\dd\Psi)\frac{1}{1-\Psi},\pi_1\bm{M_0F}\frac{1}{1-\Psi}\Big)\,.
\end{align}
This follows immediately from the identity
\begin{align}
    \pi_1\bm{M_{0,n}}\bm{F}\frac{1}{1-\Psi}=\pi_1\bm{M_{0,n}}\big(\underbrace{\pi_1\bm{F}\frac{1}{1-\Psi}\otimes\cdots\otimes \pi_1\bm{F}\frac{1}{1-\Psi}}_n\big)\,,
\end{align}
and the cyclicity of $M_{0,n}$.
Inserting the identity $\bm{I}=\bm{FF}^{-1}$ in front of $\bm{M_0}$, we obtain
\begin{align}
    \dd S_0=-\omega\Big(\pi_1\bm{F}\bm{D}(\dd\Psi)\frac{1}{1-\Psi},\pi_1\bm{F}\bm{F}^{-1}\bm{M_0F}\frac{1}{1-\Psi}\Big)\,.
\end{align}
Here, we utilize the following identity:
\begin{align}
    \bm{D}\Big(\pi_1\bm{F}^{-1}\bm{M_0F}\frac{1}{1-\Psi}\Big)\frac{1}{1-\Psi}=\bm{F}^{-1}\bm{M_0F}\frac{1}{1-\Psi}\,.\label{eq:identity}
\end{align}
The identity is easily verified by inspecting the $\pi_n$ component.
Applying $\pi_n$ to the left-hand side, and utilizing the definition of the coderivation $\bm{D}(\cdot)$ \eqref{eq:def of D} and the definition of the group-like element $\frac{1}{1-\Psi}$ \eqref{eq:def of group-like element}, the expression is written as
\begin{align}
    \pi_n\bm{D}\Big(\pi_1\bm{F}^{-1}\bm{M_0F}\frac{1}{1-\Psi}\Big)\frac{1}{1-\Psi}=\sum_{k=0}^{n-1}\Psi^{\otimes k}\otimes \pi_1\bm{F}^{-1}\bm{M_0F}\frac{1}{1-\Psi} \otimes \Psi^{\otimes n-k-1}\,.
\end{align}
On the other hand, since $\bm{F}^{-1}\bm{M_0F}$ is a coderivation, applying $\pi_n$ to the right-hand side of \eqref{eq:identity} yields
\begin{align}
    \pi_n\bm{F}^{-1}\bm{M_0F}\frac{1}{1-\Psi}=\sum_{k=0}^{n-1}\Psi^{\otimes k}\otimes \pi_1\bm{F}^{-1}\bm{M_0F}\frac{1}{1-\Psi} \otimes \Psi^{\otimes n-k-1}\,.
\end{align}
Consequently, $\dd S_0$ can be written in the following form:
\begin{align}
    \dd S_0=-\omega\Big(\pi_1\bm{F}\bm{D}(\dd\Psi)\frac{1}{1-\Psi},\pi_1\bm{F}\bm{D}\big(\pi_1\bm{F}^{-1}\bm{M_0F}\frac{1}{1-\Psi}\big)\frac{1}{1-\Psi}\Big)\,.
\end{align}
We now define a new symplectic form $\omega_{\Psi}$ by:
\begin{align}
    \omega_{\Psi}(A,B):=\omega\Big(\pi_1\bm{F}\bm{D}(A)\frac{1}{1-\Psi},\pi_1\bm{F}\bm{D}(B)\frac{1}{1-\Psi}\Big)\,,
\end{align}
for $A,B\in\mathcal{H}$.
This new symplectic form satisfies the graded anti-symmetric property.
With this, the differential of the action $\dd S_0$ becomes
\begin{align}
    \dd S_0=-\omega_{\Psi}\Big(\dd\Psi,\pi_1\bm{F}^{-1}\bm{M_0F}\frac{1}{1-\Psi}\Big)\,.\label{eq:dS0}
\end{align}
Here, $\bm{F}^{-1}\bm{M_0F}$ is a coderivation satisfying the $A_\infty$ relations.
Thus, it defines the $A_\infty$ structure of the BSFT-like action.
It is important to note that this $A_\infty$ algebra is not cyclic with respect to the new symplectic form $\omega_\Psi$. This is why the BSFT-like action obtained by integrating $\dd S_0$ does not take the Maurer–Cartan action form.

From this expression, we see that the equation of motion derived from the BSFT-like action is given by:
\begin{align}
    \pi_1\bm{F}^{-1}\bm{M_0F}\frac{1}{1-\Psi}=0\,.\label{eq:eom for BSFT-like}
\end{align}
As mentioned earlier, this equation is equivalent to the original equation of motion: $\pi_1\bm{M_0}\frac{1}{1-\Phi}=0$.
This confirms that the BSFT-like action correctly encodes the dynamics of the original Maurer-Cartan action \eqref{eq:original action}.

The differential of the BSFT action was originally given in \cite{Witten:1992qy} as
\begin{align}
\mathrm{d}S_{\mathrm{BSFT}}=\oint\dd\theta_1\oint\dd\theta_2\big\langle\mathrm{d}\big(cV(\theta_1)\big)\{Q_{\mathrm{B}},cV(\theta_2)\}\,e^{\oint\dd\theta \,V(\theta)}\big\rangle_{I_{\mathrm{CFT}}}\,,\label{eq:naiveBSFT}
\end{align}
where $\langle\cdot\rangle_{I_{\mathrm{CFT}}}$ denotes the expectation value with respect to a fixed bulk CFT action $I_{\mathrm{CFT}}$, $V$ is a local boundary operator, and $Q_\mathrm{B}$ is the worldsheet BRST operator.
By introducing the perturbed QFT action
\begin{align}
    I_{\mathrm{QFT}}=I_{\mathrm{CFT}}-\oint\dd \theta \,V(\theta)\,,
\end{align}
the differential of the BSFT action can be recast as
\begin{align}
    \mathrm{d}S_{\mathrm{BSFT}}=\oint\dd\theta_1\oint\dd\theta_2\big\langle\mathrm{d}\big(cV(\theta_1)\big)\{Q_{\mathrm{B}},cV(\theta_2)\}\,\big\rangle_{I_{\mathrm{QFT}}}\,,
\end{align}
where the expectation value is now taken with respect to $I_{\mathrm{QFT}}$.
Naively, the expression appears to vanish when $\{Q_{\mathrm{B}},cV\}=0$,  but this is not the case.
The expression \eqref{eq:naiveBSFT} involves contact divergences between boundary insertions and thus requires an appropriate cutoff and renormalization.
As a result of this procedure, $\{Q_{\mathrm{B}},cV\}$ acquires nonlinear corrections.
Incorporating these corrections and with appropriate regularization, the BSFT action can be written as
\begin{align}
    \dd S_{\mathrm{BSFT}}=\oint\dd\theta_1\oint\dd\theta_2\Big\langle\dd \big(cV(\theta_1)\big)\,\{\mathcal{Q},cV(\theta_2)\}\Big\rangle_{I_\mathrm{QFT}}\,,
\end{align}
where $\mathcal{Q}$ is a nilpotent operator that is no longer linear:
\begin{align}
    \{\mathcal{Q},cV\}=\{Q_\mathrm{B},cV\}+\mathcal{O}(V^2)\,.
\end{align}
In particular, when $V$ is a marginal operator such that $\{Q_{\mathrm{B}},cV\}=0$, the higher-order terms  are determined through amplitudes \cite{Chiaffrino:2018jfy}:
\begin{align}
   \big\langle \dd \big(cV(\pi)\big)\,\{\mathcal{Q},cV(0)\}\big\rangle_{I_{\mathrm{QFT}}}=\sum_{n}\mathcal{V}^{n+1}\big(\dd V,\underbrace{V,\dots,V}_n\big)\,,\label{eq:Q is amp}
\end{align}
where, $\mathcal{V}^n$ denotes $n$-point string amplitude.

Upon comparison with the BSFT-like action, one finds the following structural correspondences:
\begin{itemize}
    \item The string field $\Psi$ corresponds to the boundary operator $cV$.
    \item The symplectic form $\omega_{\Psi}$ corresponds to the two-point expectation value 
    \begin{align}
        \oint\dd\theta_1\oint\dd\theta_2\langle V_1(\theta_1)V_2(\theta_2)\rangle_{I_\mathrm{QFT}}\,.
    \end{align}
    \item The nilpotent coderivation $\bm{F}^{-1}\bm{M_0F}$ corresponds to the operator $\mathcal{Q}$.
\end{itemize}
In particular, the coderivation $\bm{F}^{-1}\bm{M_0F}$ acts as
\begin{align}
    \pi_1\bm{F}^{-1}\bm{M_0F}\frac{1}{1-\Psi}=Q\Psi+\pi_1\bm{Pm_0F}\frac{1}{1-\Psi}\,.
\end{align}
Here, the first term corresponds to the linear part of $\mathcal{Q}$, while the second term encodes the non-linear part arising from contact terms.
Notably,  when $\Psi$ belongs to $Q$-cohomology, the first term vanishes, and the second term reduces to the minimal model.
This behavior is fully consistent with \eqref{eq:Q is amp}.

With the $A_\infty$ structure clarified, the amplitudes generated by the BSFT-like action can be computed by constructing its minimal model via homological perturbation theory.
The non-linear part of $\bm{F}^{-1}\bm{M_0F}$ is given by $\bm{\mathcal{D}}\big(\pi_1\bm{Pm_0F}\big)$, and  the minimal model is given by
\begin{align}
    \bm{P}\bm{\mathcal{D}}\big(\pi_1\bm{Pm_0F}\big)\frac{1}{\bm{I}+\bm{h}\bm{\mathcal{D}}\big(\pi_1\bm{Pm_0F}\big)}\bm{P}\,.
\end{align}
Since $hP=0$, the term $\bm{h}\bm{\mathcal{D}}\big(\pi_1\bm{Pm_0F}\big)$ vanishes in this expression, allowing the formula to simplify to
\begin{align}
    \bm{P}\bm{\mathcal{D}}\big(\pi_1\bm{Pm_0F}\big)\frac{1}{\bm{I}+\bm{h}\bm{\mathcal{D}}\big(\pi_1\bm{Pm_0F}\big)}\bm{P}=\bm{Pm_0FP}.
\end{align}
Furthermore, by utilizing \eqref{eq:on-shell F}, we see that this expression is precisely the minimal model $\bm{Pm_0f_0P}$ of the original $A_\infty$ algebra.
The resulting on-shell amplitude takes the form:
\begin{align}
    &\omega_\Psi\Big(\Psi_1,\pi_1\bm{Pm_0f_0P}\pi_n\big(\Psi_2,\dots,\Psi_{n+1}\big)\Big)\nonumber\\
    &=\omega\Big(\pi_1\bm{F}\bm{D}(\Psi_1)\frac{1}{1-\Psi},\pi_1\bm{F}\bm{D}\big(\pi_1\bm{Pm_0f_0P}\pi_n(\Psi_2,\dots,\Psi_{n+1})\big)\frac{1}{1-\Psi}\Big)\,.
\end{align}
By using $h^2=Ph=0$, this reduces to:
\begin{align}
    \omega\Big(\Psi_1,\pi_1\bm{Pm_0f_0P}\pi_n(\Psi_2,\dots,\Psi_{n+1})\Big)\,,
\end{align}
which precisely reproduces the on-shell amplitude derived from the original action \eqref{eq:original action}.
This result is also consistent with \cite{Chiaffrino:2018jfy},
where the authors applied the Poincaré–Dulac theorem to show that the BSFT vertices map $Q$-cohomology inputs to a $Q$-cohomology output. 
In our construction, however, the output is automatically constrained to lie in the $Q$-cohomology without requiring such a procedure.

\subsection{BSFT vs. BSFT-like: action forms}

In this subsection, we compare the structure of the BSFT action and the BSFT-like action.
The BSFT action \eqref{eq:BSFT action} is given by the disk partition function of the worldsheet QFT, $Z_{D_2}$, minus the $\beta$-function—which plays the role of the equation of motion—multiplied by the derivative of the first term.
Similarly, the BSFT-like action \eqref{eq:BSFT-like action} has a comparable structure.
This can be seen as follows.
The second term of the BSFT-like action,
\begin{align}
    \omega\Big((\mathbb{I}-P)\Psi,\pi_1\bm{W}\frac{1}{1-\Psi}\Big)\,,
\end{align}
can be rewritten as
\begin{align}
    \omega\Big((\mathbb{I}-P)\Psi,\pi_1\bm{W}\frac{1}{1-\Psi}\Big)=\sum_{a}(-1)^{\deg(e_a)}\omega\Big((\mathbb{I}-P)\Psi,\widetilde{e}^a\Big)\,\omega\Big(e_a,\pi_1\bm{W}\frac{1}{1-\Psi}\Big)\,,\label{eq:BSFT-like action second term factorize}
\end{align}
where we used \eqref{eq:complete}.
The factor $\omega\big((\mathbb{I}-P)\Psi,\widetilde{e}^a\big)$ is proportional to the equation of motion.
This becomes clear upon rewriting it as
\begin{align}
    \omega\big((\mathbb{I}-P)\Psi,\widetilde{e}^a\big)=\omega\big(\pi_1\bm{hF}^{-1}\bm{M_0F}\frac{1}{1-\Psi},\widetilde{e}^a\big)\,,
\end{align}
where we used  \eqref{eq:field redef Psi} and the identity $(\mathbb{I}-P)h=h$.

Furthermore, within this factor, elements $\widetilde{e}^a$ belonging to $\mathcal{H}^{(hQ)}$ provide no contribution.
This implies that the remaining factor
\begin{align}
    (-1)^{\deg(e_a)}\omega\Big(e_a,\pi_1\bm{W}\frac{1}{1-\Psi}\Big)\,,\label{eq:pdv first term}
\end{align}
receives no contribution from elements $e_a$ belonging to $\mathcal{H}^{(Qh)}$.
Under this restriction, $e_a$ can be regarded as $\pdv{}{\psi^a}\Psi$.
Given that $\bm{W}$ is cyclic, this factor can be written as
\begin{align}
    \pdv{}{\psi^a}\sum_{n=1}^\infty\frac{1}{n+1}\omega\Big(\Psi,\pi_1\bm{W}\pi_n\frac{1}{1-\Psi}\Big)\,.
\end{align}
Therefore, the second term of the BSFT-like action \eqref{eq:BSFT-like action second term factorize} can be written as:
\begin{align}
    \sum_{\widetilde{e}^a\in\mathcal{H}^{(P)}\oplus\mathcal{H}^{(Qh)}}\omega\Big(\pi_1\bm{hF}^{-1}\bm{M_0F}\frac{1}{1-\Psi},\widetilde{e}^a\Big)\,\pdv{}{\psi^a}\sum_{n=1}^\infty\frac{1}{n+1}\omega\Big(\Psi,\pi_1\bm{W}\pi_n\frac{1}{1-\Psi}\Big)\,.
\end{align}
This structure clearly matches the second term of the BSFT action, $-\sum_a\beta^a\pdv{}{\psi^a}Z_{D_2}$.
Specifically, the derived term also carries the necessary overall minus sign and is proportional to the derivative of the first term of the action,
\begin{align}
    -\sum_{n=1}^\infty \frac{1}{n+1}\omega\Big(\Psi,\pi_1\bm{W}\pi_n\frac{1}{1-\Psi}\Big)\,,\label{eq:first term of BSFT-like action}
\end{align}
multiplied by a term proportional to the equation of motion.
Note, however, that although this term is proportional to the equation of motion, it vanishes under a weaker condition, $\Psi=P\Psi$, which is a necessary condition for the solution of the equation of motion.
In contrast, in BSFT, the equation of motion—given by the $\beta$ function—does not vanish unless the boundary operator is not only marginal but exactly marginal.
This marks a difference between the two actions.

We now turn to a comparison of the first term of the two actions.
As discussed earlier, the term 
\begin{align}
    -\omega\Big(\Psi,\pi_1\bm{W}\pi_n\frac{1}{1-\Psi}\Big)\,,
\end{align}
can be interpreted as the $n+1$-point off-shell amplitude.
Therefore, the first term of the BSFT-like action \eqref{eq:first term of BSFT-like action} represents a sum of off-shell amplitudes, where the $n$-point off-shell amplitude is weighted by a factor of $\frac{1}{n}$.
On the other hand, the first term of the BSFT action, $Z_{D_2}$,
is given by
\begin{align}
    Z_{D_2}=\int_{D_2}\mathcal{D}X\,e^{-I_{\mathrm{CFT}}+\oint\dd \theta\,V(\theta)}\,,
\end{align}
which can be formally expanded around the fixed bulk CFT as
\begin{align}
    Z_{D_2}=\sum_{n=0}^\infty \frac{1}{n!}\Big\langle \oint\dd\theta_1\,V(\theta_1)\oint\dd\theta_2\,V(\theta_2)\cdots \oint\dd\theta_n\,V(\theta_n)\Big\rangle_{I_{\mathrm{CFT}}}\,.\label{eq:formal def of Z}
\end{align}
Rewriting this expression so that $\theta_1,\theta_2,\cdots,\theta_n$
are ordered clockwise along the boundary of the disk, we obtain $(n-1)!$ identical terms. Thus, the expression can be rewritten as
\begin{align}
    Z_{D_2}=\sum_{n=0}^\infty \frac{1}{n}\Big\langle \oint_{\theta_1\leq\theta_2\leq\cdots\leq\theta_n}\dd\theta_1 \dd\theta_2\cdots\dd\theta_n\,V(\theta_1)V(\theta_2)\cdots V(\theta_n)\Big\rangle_{I_{\mathrm{CFT}}}\,.
\end{align}
Here,the integration over $\theta_i$ is performed while preserving the cyclic ordering.
This can be interpreted as a
sum of off-shell string amplitudes, just like the first term of the BSFT-like action \eqref{eq:first term of BSFT-like action}, and the weighting factor $\frac{1}{n}$ matches.
The $n=0$ and $n=1$ terms are absent in the first term of the BSFT-like action.
However, this is not significant.
The $n=0$ term is a constant and can be ignored.
The $n=1$ term is non-zero only when the conformal dimension of $V$ is zero, but in that case, it cancels with the linear term of $V$ in the second term of the BSFT action. 

However, this correspondence remains formal. 
The expression \eqref{eq:formal def of Z} is ill-defined due to contact divergences, and in practice, an appropriate cutoff and renormalization are required. 
As a result, the first term of the BSFT action, $Z_{D_2}$, deviates slightly from a sum of off-shell amplitudes. 
This also marks a distinction from the BSFT-like action.
Nevertheless, the on-shell behavior of the full action agrees exactly.
In that case, the second term of the BSFT-like action vanishes, and the first term reduces to the on-shell amplitudes; whereas in BSFT, the second term does not vanish, but the full action still reduces to the on-shell amplitudes  \cite{Chiaffrino:2018jfy}.
While the first and second terms of the BSFT action and the BSFT-like action differ slightly—due to the regularization procedure—these differences cancel out in the full action, at least at the on-shell level.
This suggests that the BSFT-like action may provide a natural regularization of BSFT.
To verify this, it is important for future work to investigate whether the BSFT-like action—with appropriately chosen string products—can be matched with the BSFT action with a suitable cutoff and added counterterms.

Moreover, our analysis has been carried out entirely within perturbation theory. 
Consequently, we are not able to address the non-perturbative relationship between the Maurer–Cartan action and the BSFT action in this framework. 
Extending our construction beyond the perturbative regime remains an important open problem.
Further investigations are necessary to determine whether the field redefinition remains valid non-perturbatively and to clarify the conditions under which the BSFT action can be fully reconstructed from the Maurer–Cartan action. Such studies could deepen our understanding of the connection between the algebraic structure of string field theory and BSFT.

\section{Conclusion}\label{sec:conclusion}

In this work, we have proposed an approach to constructing the BSFT-like action by employing field redefinitions based on $A_\infty$ algebraic structures.
Specifically, we introduced a cohomomorphism that maps the Maurer–Cartan action of a given $A_\infty$ algebra to an expression closely resembling the BSFT formulation. 
This transformation provides a systematic framework for deriving off-shell amplitudes while maintaining structural similarities with BSFT.

A key insight from our construction is that, if the original Maurer–Cartan action is appropriately chosen, the BSFT-like action is expected to reproduce the BSFT action itself. This suggests a deep connection between the algebraic structures of string field theory and the BSFT framework.
In particular, our formulation highlights how field redefinitions within homotopy algebras naturally give rise to BSFT-like structures.

An important direction for future work is to investigate whether this approach indeed provides a proper regularization of BSFT.
While our analysis has focused on the perturbative regime, another important direction is to explore the non-perturbative aspects of this framework. Understanding whether the field redefinition remains well-defined beyond perturbation theory could provide further insights into the fundamental structure of string field theory.

Throughout this paper, we have consistently neglected the on-shell two-point amplitude. Addressing this issue remains an open problem. A promising direction is suggested in \cite{Alfonsi:2024utl}, where the inclusion of boundary terms plays a key role. It will be important to understand how our framework is modified when such boundary contributions are taken into account.

Furthermore, our approach is expected to extend naturally to closed string field theory formulated in terms of $L_\infty$ algebras, as well as to open-closed string field theory described by open-closed homotopy algebras.
In particular, for closed strings, the resulting formulation is anticipated to be related to the C-function, as discussed in \cite{Ahmadain:2024hdp}.

\bigskip

\noindent
{\normalfont \bfseries \large Acknowledgments}

\noindent I would like to thank Keisuke Konosu, Yuji Okawa, and Ivo Sachs for helpful discussions.
Part of this work was presented at the \textit{Workshop on String Field Theory and Related Aspects} held in March of 2025 in Trieste.
This work is supported by JSPS KAKENHI Grant Nos. JP23KJ1311.


\medskip



\begin{thebibliography}{999}

\bibitem{Stasheff:I}
  J.~D.~Stasheff,
  ``Homotopy associativity of $H$-spaces. I,''
  Trans. of the Amer. Math. Soc. {\bf 108}, 275 (1963).

\bibitem{Stasheff:II}
  J.~D.~Stasheff,   ``Homotopy associativity of $H$-spaces. II,''
  Trans. of the Amer. Math. Soc. {\bf 108}, 293 (1963).

\bibitem{Getzler-Jones}
  E.~Getzler and J.~D.~S.~Jones,
  ``$A_\infty$-algebras and the cyclic bar complex,''
  Illinois~J.~Math {\bf 34}, 256 (1990).

\bibitem{Markl}
  M.~Markl,
  ``A cohomology theory for $A (m)$-algebras and applications,''
  J. Pure Appl. Algebra {\bf 83}, 141 (1992).

\bibitem{Penkava:1994mu}
  M.~Penkava and A.~S.~Schwarz,
  ``$A_\infty$ algebras and the cohomology of moduli spaces,''
  hep-th/9408064.

\bibitem{Gaberdiel:1997ia}
  M.~R.~Gaberdiel and B.~Zwiebach,
  ``Tensor constructions of open string theories. 1: Foundations,''
  Nucl. Phys. {\bf B505}, 569 (1997)
  [hep-th/9705038].


\bibitem{Zwiebach:1992ie}
B.~Zwiebach,
``Closed string field theory: Quantum action and the Batalin-Vilkovisky master equation,''
Nucl. Phys. B \textbf{390}, 33-152 (1993)
[arXiv:hep-th/9206084 [hep-th]].

\bibitem{Markl:1997bj}
M.~Markl,
``Loop homotopy algebras in closed string field theory,''
Commun. Math. Phys. \textbf{221}, 367-384 (2001)
[arXiv:hep-th/9711045 [hep-th]].

\bibitem{Maccaferri:2022yzy}
C.~Maccaferri and J.~Vo\v{s}mera,
``The classical cosmological constant of open-closed string field theory,''
JHEP \textbf{10}, 173 (2022)
[arXiv:2208.00410 [hep-th]].

\bibitem{Maccaferri:2023gcg}
C.~Maccaferri, A.~Ruffino and J.~Vo\v{s}mera,
``The nilpotent structure of open-closed string field theory,''
JHEP \textbf{08}, 145 (2023)
[arXiv:2305.02843 [hep-th]].

\bibitem{Kajiura:2004xu}
H.~Kajiura and J.~Stasheff,
``Homotopy algebras inspired by classical open-closed string field theory,''
Commun. Math. Phys. \textbf{263}, 553-581 (2006)
[arXiv:math/0410291 [math.QA]].

\bibitem{Kajiura:2006mt}
H.~Kajiura and J.~Stasheff,
``Homotopy algebra of open-closed strings,''
Geom. Topol. Monographs \textbf{13}, 229-259 (2008)
[arXiv:hep-th/0606283 [hep-th]].

\bibitem{Erler:2013xta}
T.~Erler, S.~Konopka and I.~Sachs,
``Resolving Witten`s superstring field theory,''
JHEP \textbf{04}, 150 (2014)
[arXiv:1312.2948 [hep-th]].

\bibitem{Erler:2016ybs}
T.~Erler, Y.~Okawa and T.~Takezaki,
``Complete Action for Open Superstring Field Theory with Cyclic $A_\infty$ Structure,''
JHEP \textbf{08}, 012 (2016)
[arXiv:1602.02582 [hep-th]].

\bibitem{Kunitomo:2019glq}
H.~Kunitomo and T.~Sugimoto,
``Heterotic string field theory with cyclic $L_\infty$ structure,''
PTEP \textbf{2019}, no.6, 063B02 (2019)
[arXiv:1902.02991 [hep-th]].

\bibitem{Kunitomo:2022qqp}
H.~Kunitomo,
``Open-closed homotopy algebra in superstring field theory,''
PTEP \textbf{2022}, no.9, 093B07 (2022)
[arXiv:2204.01249 [hep-th]].

\bibitem{Erbin:2020eyc}
H.~Erbin, C.~Maccaferri, M.~Schnabl and J.~Vo\v{s}mera,
``Classical algebraic structures in string theory effective actions,''
JHEP \textbf{11}, 123 (2020)
[arXiv:2006.16270 [hep-th]].

\bibitem{Koyama:2020qfb}
D.~Koyama, Y.~Okawa and N.~Suzuki,
``Gauge-invariant operators of open bosonic string field theory in the low-energy limit,''
[arXiv:2006.16710 [hep-th]].

\bibitem{Arvanitakis:2020rrk}
A.~S.~Arvanitakis, O.~Hohm, C.~Hull and V.~Lekeu,
``Homotopy Transfer and Effective Field Theory I: Tree-level,''
Fortsch. Phys. \textbf{70}, no.2-3, 2200003 (2022)
[arXiv:2007.07942 [hep-th]].

\bibitem{Arvanitakis:2021ecw}
A.~S.~Arvanitakis, O.~Hohm, C.~Hull and V.~Lekeu,
``Homotopy Transfer and Effective Field Theory II: Strings and Double Field Theory,''
Fortsch. Phys. \textbf{70}, no.2-3, 2200004 (2022)
[arXiv:2106.08343 [hep-th]].

\bibitem{Bonezzi:2023xhn}
R.~Bonezzi, C.~Chiaffrino, F.~Diaz-Jaramillo and O.~Hohm,
``Tree-level Scattering Amplitudes via Homotopy Transfer,''
[arXiv:2312.09306 [hep-th]].

\bibitem{Konopka:2015tta}
S.~Konopka,
``The S-Matrix of superstring field theory,''
JHEP \textbf{11}, 187 (2015)
[arXiv:1507.08250 [hep-th]].

\bibitem{Kunitomo:2020xrl}
H.~Kunitomo,
``Tree-level S-matrix of superstring field theory with homotopy algebra structure,''
JHEP \textbf{03}, 193 (2021)
[arXiv:2011.11975 [hep-th]].

\bibitem{Nutzi:2018vkl}
A.~N\"utzi and M.~Reiterer,
``Amplitudes in YM and GR as a Minimal Model and Recursive Characterization,''
Commun. Math. Phys. \textbf{392}, no.2, 427-482 (2022)
[arXiv:1812.06454 [math-ph]].

\bibitem{Arvanitakis:2019ald}
A.~S.~Arvanitakis,
``The $L_\infty$-algebra of the S-matrix,''
JHEP \textbf{07}, 115 (2019)
[arXiv:1903.05643 [hep-th]].

\bibitem{Macrelli:2019afx}
T.~Macrelli, C.~S\"amann and M.~Wolf,
``Scattering amplitude recursion relations in Batalin-Vilkovisky\textendash{}quantizable theories,''
Phys. Rev. D \textbf{100}, no.4, 045017 (2019)
[arXiv:1903.05713 [hep-th]].

\bibitem{Jurco:2019yfd}
B.~Jur\v{c}o, T.~Macrelli, C.~S\"amann and M.~Wolf,
``Loop Amplitudes and Quantum Homotopy Algebras,''
JHEP \textbf{07}, 003 (2020)
[arXiv:1912.06695 [hep-th]].

\bibitem{Okawa:2022sjf}
Y.~Okawa,
``Correlation functions of scalar field theories from homotopy algebras,''
JHEP \textbf{05}, 040 (2024)
[arXiv:2203.05366 [hep-th]].

\bibitem{Konosu:2023pal}
K.~Konosu and Y.~Okawa,
``Correlation functions involving Dirac fields from homotopy algebras I: the free theory,''
[arXiv:2305.11634 [hep-th]].

\bibitem{Konosu:2023rkm}
K.~Konosu,
``Correlation Functions Involving Dirac Fields from Homotopy Algebras II: The Interacting Theory,''
PTEP \textbf{2024}, no.9, 093B01 (2024)
[arXiv:2305.13103 [hep-th]].

\bibitem{Konosu:2024dpo}
K.~Konosu and J.~Totsuka-Yoshinaka,
``Noether\textquoteright{}s theorem and Ward-Takahashi identities from homotopy algebras,''
JHEP \textbf{09}, 048 (2024)
[arXiv:2405.09243 [hep-th]].

\bibitem{Konosu:2025bnz}
K.~Konosu, Y.~Okawa, S.~Shibuya and J.~Totsuka-Yoshinaka,
``The LSZ reduction formula from homotopy algebras,''
[arXiv:2504.08653 [hep-th]].

\bibitem{Konosu:2024zrq}
K.~Konosu and Y.~Okawa,
``Nonperturbative correlation functions from homotopy algebras,''
JHEP \textbf{01} (2025), 152
[arXiv:2405.10935 [hep-th]].

\bibitem{Erler:2020beb}
T.~Erler and H.~Matsunaga,
``Mapping between Witten and lightcone string field theories,''
JHEP \textbf{11}, 208 (2021)
[arXiv:2012.09521 [hep-th]].

\bibitem{Ando:2024pmr}
Y.~Ando, R.~Fujii, H.~Kunitomo and J.~Totsuka-Yoshinaka,
``A consistent light-cone-gauge superstring field theory,''
[arXiv:2411.19570 [hep-th]].

\bibitem{JalaliFarahani:2023sfq}
M.~Jalali Farahani, C.~Saemann and M.~Wolf,
``Field theory equivalences as spans of -algebras,''
J. Phys. A \textbf{57} (2024) no.28, 285208
[arXiv:2305.05473 [hep-th]].


\bibitem{Witten:1992qy}
E.~Witten,
``On background independent open string field theory,''
Phys. Rev. D \textbf{46} (1992), 5467-5473
[arXiv:hep-th/9208027 [hep-th]].

\bibitem{Witten:1992cr}
E.~Witten,
``Some computations in background independent off-shell string theory,''
Phys. Rev. D \textbf{47} (1993), 3405-3410
doi:10.1103/PhysRevD.47.3405
[arXiv:hep-th/9210065 [hep-th]].

\bibitem{Shatashvili:1993kk}
S.~L.~Shatashvili,
``Comment on the background independent open string theory,''
Phys. Lett. B \textbf{311} (1993), 83-86
[arXiv:hep-th/9303143 [hep-th]].

\bibitem{Shatashvili:1993ps}
S.~L.~Shatashvili,
``On the problems with background independence in string theory,''
Alg. Anal. \textbf{6} (1994), 215-226
[arXiv:hep-th/9311177 [hep-th]].

\bibitem{Li:1993za}
K.~Li and E.~Witten,
``Role of short distance behavior in off-shell open string field theory,''
Phys. Rev. D \textbf{48} (1993), 853-860
[arXiv:hep-th/9303067 [hep-th]].

\bibitem{Gerasimov:2000zp}
A.~A.~Gerasimov and S.~L.~Shatashvili,
``On exact tachyon potential in open string field theory,''
JHEP \textbf{10} (2000), 034
[arXiv:hep-th/0009103 [hep-th]].

\bibitem{Kutasov:2000qp}
D.~Kutasov, M.~Marino and G.~W.~Moore,
``Some exact results on tachyon condensation in string field theory,''
JHEP \textbf{10} (2000), 045
[arXiv:hep-th/0009148 [hep-th]].

\bibitem{Kutasov:2000aq}
D.~Kutasov, M.~Marino and G.~W.~Moore,
``Remarks on tachyon condensation in superstring field theory,''
[arXiv:hep-th/0010108 [hep-th]].

\bibitem{Takayanagi:2000rz}
T.~Takayanagi, S.~Terashima and T.~Uesugi,
``Brane - anti-brane action from boundary string field theory,''
JHEP \textbf{03} (2001), 019
[arXiv:hep-th/0012210 [hep-th]].

\bibitem{Hashimoto:2015iha}
K.~Hashimoto, S.~Sugishita and S.~Terashima,
``Ramond-Ramond couplings of D-branes,''
JHEP \textbf{03} (2015), 077
[arXiv:1501.00995 [hep-th]].

\bibitem{Costello:2016vjw}
K.~Costello and O.~Gwilliam,
``Factorization Algebras in Quantum Field Theory: Volume 1,''
Cambridge University Press (2016).

\bibitem{Costello:2021jvx}
K.~Costello and O.~Gwilliam,
``Factorization Algebras in Quantum Field Theory: Volume 2,''
Cambridge University Press (2021).

\bibitem{Erbin:2019uiz}
H.~Erbin, J.~Maldacena and D.~Skliros,
``Two-Point String Amplitudes,''
JHEP \textbf{07} (2019), 139
[arXiv:1906.06051 [hep-th]].

\bibitem{Alfonsi:2024utl}
L.~Alfonsi, L.~Borsten, H.~Kim, M.~Wolf and C.~A.~S.~Young,
``Full S-matrices and Witten diagrams with (relative) L-infinity algebras,''
[arXiv:2412.16106 [hep-th]].

\bibitem{Chiaffrino:2018jfy}
C.~Chiaffrino and I.~Sachs,
``Classical open string amplitudes from boundary string field theory,''
JHEP \textbf{06} (2019), 086
[arXiv:1805.07084 [hep-th]].


\bibitem{Ahmadain:2024hdp}
A.~Ahmadain, A.~Frenkel and A.~C.~Wall,
``A Background-Independent Closed String Action at Tree Level,''
[arXiv:2410.11938 [hep-th]].






\end{thebibliography}
\end{document}